\documentclass[useAMS,usenatbib]{mn2e}
\usepackage {graphicx}

\title[Detecting the effect of Globular Cluster impacts on the disk of the Milky Way]{Detecting the effect of Globular Cluster impacts on the disk of the Milky Way}

\author[D. Vande Putte and Mark Cropper]{D. Vande Putte\thanks{E-mail:
dwvp@mssl.ucl.ac.uk (DVP)} and Mark Cropper\thanks{E-mail:
msc@mssl.ucl.ac.uk (MC)}
\\
University College London, Mullard Space Science Laboratory, Holmbury St Mary, Dorking, RH5 6NT, UK\\}

\begin{document}

\date{Accepted yyyy mm dd. Received  yyyy mm dd; in original form  yyyy mm dd}

\pagerange{\pageref{firstpage}--\pageref{lastpage}} \pubyear{2002}

\maketitle

\label{firstpage}

\begin{abstract}

The crossing of the Galactic disk by a Globular Cluster could produce star formation due to gravitational focussing or compression of disk material.  We report on simulations of the effect on disk material which reveal that the crossing can sometimes cause local gravitational focussing of disk material.  We also present the salient points of a little-known paper by Levy (2000), that shows that strong compression can result from the shock wave generated by GC disk crossing.  The main thrust of our paper is a search for remnants of disk crossings by Globular Clusters.  Using the gravitational potential of the Galaxy to locate the position of the most recent crossings of a subset of fifty-four Globular Clusters reveals that systematic errors and uncertainties in initial conditions limit the scope for unequivocal identification.  From the subset of fifty-four, six possible search sites with the best constraints are retained for further scrutiny. Three of the six potentially promising search areas in the disk are from Globular Clusters NGC 3201, 6397 and NGC 6838, for which we cannot rule out some observed star associations observed nearby as being remnants.  The three other of the six areas are too large to provide meaningful identification of remnants.  Also, a possible remnant (open cluster NGC6231) is shown not to be due to Globular Cluster impact, contrary to a previous report. In a more wide-ranging screening of one hundred and fifty-five Globular Clusters we identify which Globular Clusters are compatible with being responsible for the formation of any of the Galaxy's five most prominent Star Super Clusters. 
\end{abstract}


\section{Introduction}

One of the less-studied interactions within our Galaxy concerns the effect of the passage of a Globular Cluster (GC) through the disk, on the disk itself. On average, one Globular Cluster will cross the Galactic disk about every million years.  While 
the Globular Cluster is in the vicinity of the disk, its gravitational field 
will attract stars and gas in the disk. The result could be a density 
increase sufficient to cause star formation.  Whereas the impacts of High Velocity Clouds with the disk have been examined in some detail (Comer\'{o}n and Torra, 1994), as they cause gas compression and star formation (and are perhaps the origin of the Gould belt), the effect of GC impacts on the disk has received almost no attention.  The earliest mention of 
the effect of disk crossing by a Globular Cluster on the disk, is from 
Brosche et al. (1991) who calculated the orbits of Globular Clusters NGC362 and NGC6218, and suggested observers examine their latest crossing points, to see whether any remnants may be present.  Wallin, Higdon \& Staveley-Smith (1996) developed the idea. Their main argument is that gravitational focussing of disk material, such as gas, towards the GC trajectory would occur.  They concluded 
that large OB star associations would form, possibly some 30 Myr after the 
disk crossing, over a region of size $\sim$10 pc. Their paper 
elicited little reaction, aside from two instances.  The first was by Levy (2000), who examined compression of disk gas by the shock wave produced during GC travel through the disk.  This paper appears never to have been cited, in spite of its importance to this question; we refer to it below.  The second reaction to Wallin et al's paper was the suggestion by R. Rees and K. 
Cudworth, (reported in Wright, 2004) that the star association NGC 6231 in 
the Galactic disk is the result of such an interaction by Globular Cluster 
NGC6397, less than 5Myr ago. They reached this conclusion using positional 
and proper motion data, together with a code to calculate orbits in the 
Galactic potential. They found the point of impact of NGC6397 to correspond 
to the position of open cluster NGC6231, a finding we discuss later.

The phenomena that take place in disk crossings by GCs are akin to those encountered in minor mergers and galaxy harassment, where a large galaxy interacts with a smaller galaxy at low and high relative velocities, respectively.  Cox et al. (2006) predict that the global star formation rate decreases with increasing mass ratio, so that mergers with mass ratio$ \, > $ 20 induce hardly any additional star formation in the more massive galaxy, above its quiescent level.  In the case of GCs, the mass ratios are even larger, but relative velocities are lower and local effects in the disk may still be significant.  Indeed, minor mergers have been proposed as the origin of nuclear starburst rings in some local galaxies (e.g. Mazzuca et al. 2006).

The objective of this work is to examine whether, in principle, star formation, or at least disk material compression could occur in a GC impact, and to look for possible remnants in the Galaxy.  In section 2, we describe an approach to modelling the interaction between a GC and the disk, which establishes how compression may occur via gravitational focussing.  This includes developing models of the Galaxy and GC potentials to calculate the trajectories of  GCs and test masses in the disk.  We then note an alternate mechanism for compression, namely shock wave generation.  If gas is present, such a mechanism will be more effective at inducing star formation.  In section 3, we apply the Galactic models to locate the most recent impact areas of GCs in the disk.  Sections 4 and 5 use these location results to search for possible remnants in the Galactic plane.  Finally, section 6 provides conclusions.

\bigskip
A preliminary version of this work was reported in Cropper \& Vande Putte (2007).

\section{MODELLING}

We examine first the case of a gas-free disk, and consider the effect of a GC crossing, using numerical simulations of a series of test masses representing disk material encountering a Globular Cluster as it crosses the disk. These test 
masses are susceptible to the Galactic potential, and to the potential of the 
Globular Cluster, but not to one another.  We later consider the case where gas is present at the impact site.  First, however, we use the next two subsections to describe the methods we employ to calculate orbits in Galactic and GC potentials.

\subsection{The Galactic potentials}
We use four Galactic potentials for which straightforward analytical expressions are 
available. We denote these DA, FE, FL, and PA, as they appear in Dauphole \& 
Colin (1995), Fellhauer et al. (2006), Flynn, Sommer-Larsen \& Christensen 
(1996), and Paczynski (1990), respectively. Others have used these 
expressions in studies of galactic dynamics, for example Kalirai et al. 
(2007), Law, Johnston \& Majewski (2005), Pauli et al. (2003), and Dinescu, Girard \& van Altena (1999).

These potentials are all axisymmetric, so their expressions are simplest 
in cylindrical coordinates $R,\phi ,z$ where $R$ is the distance to the z 
axis, $\phi $ is the azimuthal angle between the x axis and the projection 
of the position vector on the xy plane, and z the distance 
above the x, y plane.

The analytical characteristics of the potentials appear in Table 1.  They consist of bulge, halo, and disk contributions described by different models, as follows.

\begin{table*}
\centering
\caption{Characteristics of selected potentials}
\begin{tabular}
{|p{69pt}|p{71pt}|p{58pt}|p{83pt}|p{50pt}|p{64pt}|}
\hline
Potential& 
Bulge& 
Halo& 
Disk& 
V$_{circ}$ at 8kpc& 
\textit{M}$_{total}$ at 100kpc \\
& 
& 
& 
& 
km s$^{ - 1}$& 
10$^{11}$ $M_{ \odot} $\\
\hline
Dauphole (DA)& 
Plummer& 
Plummer& 
Miyamoto-Nagai& 
225& 
7.9 \\
Fellhauer (FE)& 
Hernquist& 
log form& 
Miyamoto-Nagai& 
224& 
9.3 \\
Flynn (FL)& 
2xPlummer& 
log form& 
3xMiyamoto-Nagai& 
222& 
11.8 \\
Paczynski (PA)& 
Miyamoto-Nagai& 
log-atan form& 
Miyamoto-Nagai& 
220& 
8.5 \\
\hline
\end{tabular}
\end{table*}

The analytical expression for the potential from a Plummer (1911) sphere  is:

\begin{equation}
\label{eq2}
\Phi (r)\, = \, - \,{\frac{{GM_{b}} }{{{\left[ {r^{2}\, + 
\,b^2}  \right]}^{1 / 2}}}}
\end{equation}
where $r^{2}\, = \,R^{2}\, + \,z^{2}.$

The  Miyamoto-Nagai (1975) potential's analytical expression is:

\begin{equation}
\label{eq3}
\Phi (R,z)\, = \, - \,{\frac{{GM_{d}} }{{{\left\{ {R^{2}\, + \,{\left[ 
{a\, + \,(z^{2} + b^{2})^{1 / 2}} \right]}^{2}} \right\}}^{1 / 2}}}}
\end{equation}

The expression for the Hernquist (1990) potential is:

\begin{equation}
\label{eq5}
\Phi (r)\, = \, - \,{\frac{{GM} }{{r\, + \,a}}}
\end{equation}
\noindent

The logarithmic potential form is (Fellhauer et al. 2006):

\begin{equation}
\label{eq6}
\Phi(r) \, = \,{\frac{{v_{0}^{2}} }{{2}}}\,\ln (r^{2}\, + \,d^{2})
\end{equation}

The logarithmic-arctan combination potential is (Paczynski, 1990):

\begin{equation}
\label{eq9}
\Phi(r) \, = \,{\frac{{GM} }{{r_{h}} }}\,{\left[ {{\frac{{1}}{{2}}}\ln 
(1 + {\frac{{r^{2}}}{{r_{h}^{2}} }})\, + \,{\frac{{r_{h}} }{{r}}}{\rm a}\tan 
({\frac{{r}}{{r_{h}} }})} \right]}
\end{equation}

\begin{figure*}
\begin{center}
\begin{tabular}{c@{\hspace{1pc}}c}
\includegraphics[bb=23 40 476 469,height=7.5cm]{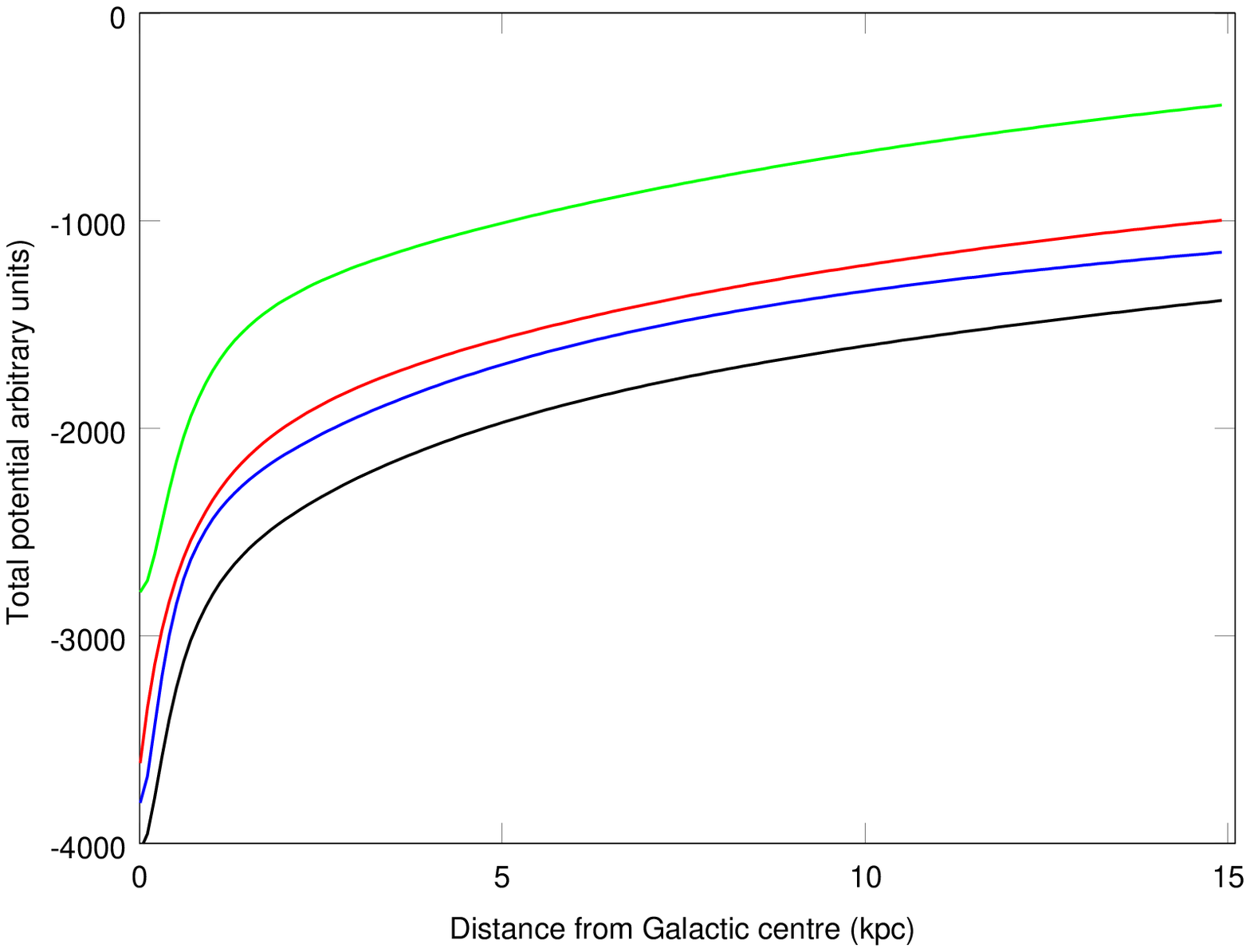} &
\includegraphics[bb=23 40 576 469,height=7.5cm]{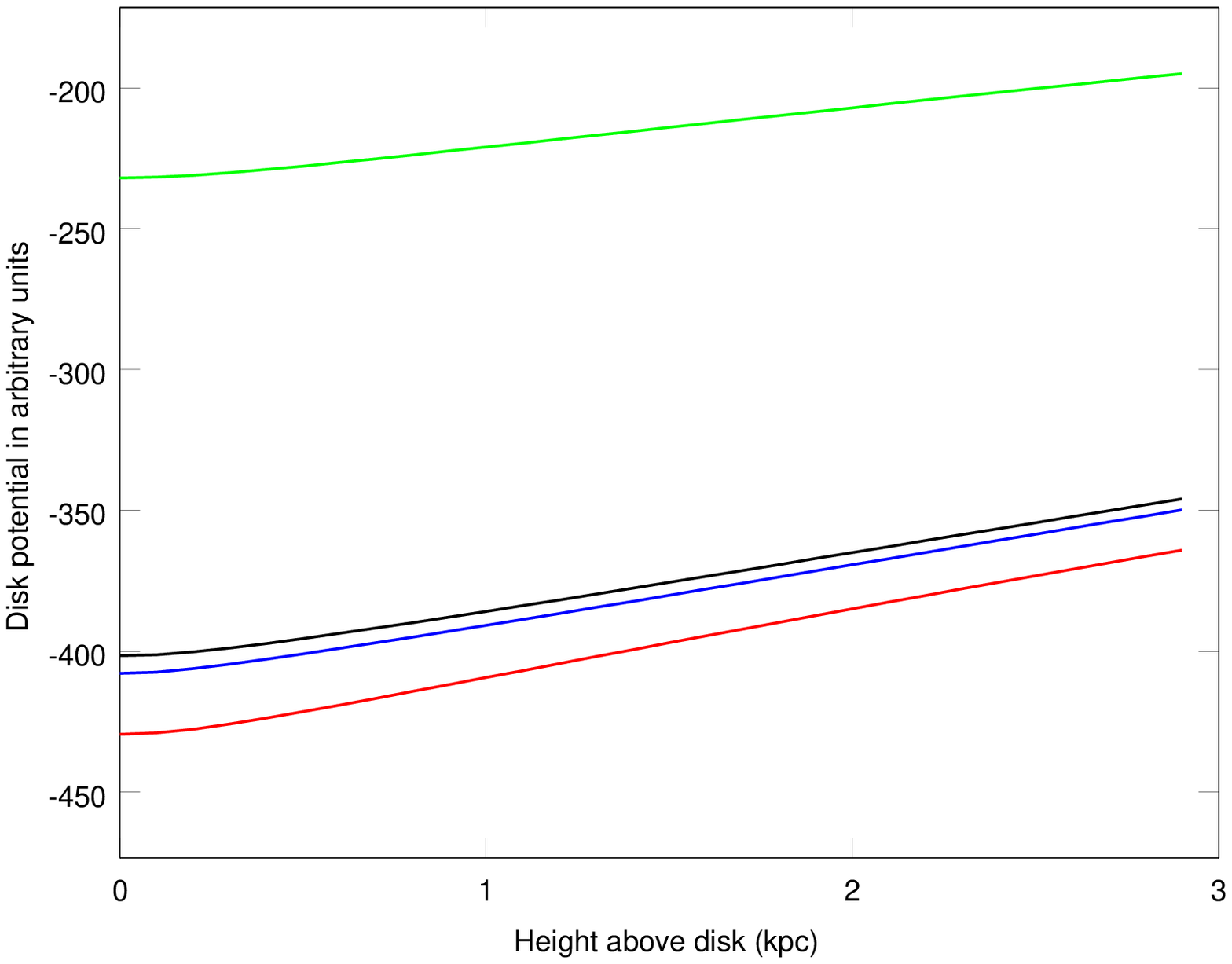} \\
\end{tabular}
\end{center}
\caption{Left panel, radial variation of potentials; right panel, vertical variation of potentials at \textit{R}=8 kpc. Potential: DA, black; FE, red; FL, 
green; PA, blue.}
\end{figure*}

\begin{table*}
\centering
\caption{Parameter values for the potentials}
\begin{tabular}{c|clcc|ccc}
\hline
  & \multicolumn{2}{c}{-----------------Bulge-----------------} & \multicolumn{1}{c}{Halo} & \multicolumn{3}{c}{----------------------------Disk----------------------------}  \\
\hline
Dauphole& 
$M$=1.3955x10$^{10}$& 
& 
$M$=6.9776x10$^{11}$& 
& 
$M$=7.9080x10$^{10}$& 
 \\
%\hline
& 
$b$=0.35& 
& 
$b$=24& 
& 
$a$=3.55& 
 \\
%\hline
& 
& 
& 
& 
& 
$b$=0.25& 
 \\
\hline
Fellhauer& 
$M$=3.4x10$^{10}$& 
& 
$v_{0}$=186 km s$^{ - 1}$& 
& 
$M$=10$^{11}$& 
 \\
%\hline
& 
$a$=0.7& 
& 
$d$=12& 
& 
$a$=6.5& 
 \\
%\hline
& 
& 
& 
& 
& 
$b$=0.26& 
 \\
\hline
Flynn& 
$M$=3x10$^{9}$& 
$M$=1.6x10$^{10}$& 
$v_{0}$=220 km s$^{ - 1}$& 
$M$=6.6x10$^{10}$& 
$M$=-2.9x10$^{10}$& 
$M$=3.3x10$^{9}$ \\
%\hline
& 
$b$=2.7& 
$b$=0.42& 
$d$=8.5& 
$a$=5.81& 
$a$=17.43& 
$a$=34.86 \\
%\hline
& 
& 
& 
& 
$b$=0.3& 
$b$=0.3& 
$b$=0.3 \\
\hline
Paczynski& 
$M$=1.12x10$^{10}$& 
& 
$M$=5x10$^{10}$& 
& 
$M$=8.07x10$^{10}$& 
 \\
%\hline
& 
$a$=0& 
& 
$r_{h}$=6.0& 
& 
$a$=3.7& 
 \\
%\hline
& 
$b$=0.0277& 
& 
& 
& 
$b$=0.2& 
 \\
\hline
\end{tabular}
\\ \S Masses (M) are in units of Solar mass, halo velocities $v_{0}$ in kms$^{{\rm -} {\rm 1}}$, and remaining parameters in kpc.  The numbers of significant figures for parameter values reflect the situation in the original papers.  The second disk mass in the FL potential is negative, but the disk and total densities are nevertheless positive everywhere.
\\
\end{table*}

Figure 1 shows a comparison between the potentials, radially along the disk, and as a function of $z$ at \textit{R} = 8kpc.
The parameter values for the different cases appear in Table 2.

All calculations are undertaken for a fixed Cartesian Galactocentric 
right-handed coordinate system. The y-axis points from the Galactic centre 
to the Sun, and the z-axis points to the North Galactic Pole (NGP). The 
x-axis therefore points in the direction of motion of the Local Standard of 
Rest (LSR). The initial position of an object results from its 
Heliocentric distance, and its celestial coordinates. The velocities of an 
object with respect to the LSR are determined from its proper motion and 
radial velocity, and from the Solar motion (U$_{\odot}$=10$\pm 
$0.4, V$_{\odot }$=5.2$\pm $0.6, W$_{\odot}$=7.2$\pm $0.4 kms$^{{\rm -} {\rm 1}}$ from Binney \& Merrifield 1998, 
p628). In the LSR's coordinate system, the U axis points towards the 
Galactic centre, the V axis points in the direction of motion of the LSR, 
and the W axis points to the NGP. We transform radial 
velocities and proper motions to motion in the LSR (U, V, W), using the formalism of Johnson \& Soderblom (1987). The final 
transformation from velocities with respect to the LSR to velocities in a 
fixed Galactocentric system uses the LSR velocity given above 
for the four potentials, with an uncertainty of $\pm $15 kms$^{{\rm - 
}{\rm 1}}$ suggested by Binney \& Tremaine (1994, p14). 

We prepared fortran95 computer codes to calculate orbits in the four potentials. We numerically integrate the 
equations of motion that arise from the potentials in cylindrical coordinates, using a fourth order 
Runge-Kutta method. The codes require input of the 
initial position and velocity. We have verified the code by comparison with detailed numerical results from another 
institute (Turku Observatory in Finland), and by comparisons with published orbits, 
and discussions with their authors.  The codes generate a series of files with the positions, 
velocities, total energy, $z$-component of angular momentum and elapsed time 
values. They also produce data on the position and timing of disk crossings, 
including the velocity at time of crossing. The user is able to specify the 
length of time and output times for the Runge-Kutta integration.  All standard routines are from the NAG library \footnote{www.nag.co.uk}.

\subsection{Addition of a moving potential}

The codes can also include a moving potential, such as that from a GC. In this case, the massive 
object experiences the potential of the Galaxy, whereas the test masses
experience the potential of the Galaxy, as well as that of the massive 
object, but not of one another. We represent the GC's potential by a Plummer expression, where the 
parameter $b$ in Eqn. (1) is derived from the mass of the Globular cluster and 
its central density \textit{$\rho $}$_{{\rm 0}{\rm ,}{\rm} }$using the expression from 
Binney \& Tremaine (Eqn. 2-47b, 1994):

\begin{equation}
\label{eq10}
\rho _{0} \, = \,{\frac{{3M}}{{4\pi b^{3}}}}{\rm .}
\end{equation}

\subsection{Gravitational focussing on a disk scale}

Wallin et al. (1996) make their case for star formation by considering phenomena at a disk scale of $\sim$ 0.1kpc, and we now investigate this in detail.

Taking a set of fifty-four GCs described in more detail in section 3.1, we calculated the relative velocity, in the Galaxy, of each GC with respect to a cloud of disk material (test masses) on a circular orbit that meets the GC at disk crossing.  We then model the cloud as 30,000 test masses placed at random in a box 0.1 kpc wide, 0.1 kpc deep, and 0.1 kpc high, with the GC impacting the box along the normal, at the relevant velocity aimed at the centre of the box.  We follow the test masses in the potential of the GC, until the GC leaves the box.  Figure 2 shows the cloud of test masses in the box viewed along the GC trajectory for NGC5139 ($\omega$ Cen), the most massive Galactic GC, albeit here for only 2,000 test masses, to avoid crowding the diagram.  Gravitational focussing towards the centre is clearly visible, and is expected to be even larger when the gravity of the test masses themselves is accounted for.

\begin{figure}
\includegraphics[bb= 0 0 400 383,width=84mm]{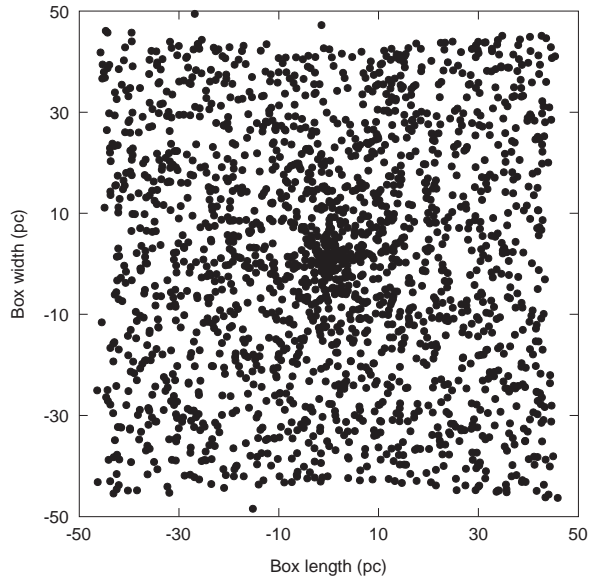}
\caption{View of box of 2,000 test masses after NGC5139  ($\omega$ Cen) traverses it}
\end{figure}

We repeated this experiment for a range of GC masses and velocities that cover the cases represented by our set of fifty-four GCs, and examined the compression as a function of GC mass/velocity, 50Myr after start of the experiment, when all GCs have exited the box.  Compression is shown in Figure 3, for three values of the parameter $b$ in the Plummer potential (Eqn.1).  Generally, compression increases with increasing mass, decreasing velocity, and is affected by the value of $b$.  For small values of mass/velocity, compression is higher for small $b$.  As mass/velocity increases, the potential gradient is steeper, and the test masses are rapidly scattered away from the GC.  For higher $b$, the gradient is smaller(softer potential) and test masses spend a longer time in the vicinity of the GC track, leading to a ring of increased compression, due to an overlap of test masses moving to, and from the GC.

\begin{figure}
%\begin{center}
\begin{tabular}{c@{\hspace{1pc}}c}
\includegraphics[bb= 0 0 495 390,width=69mm]{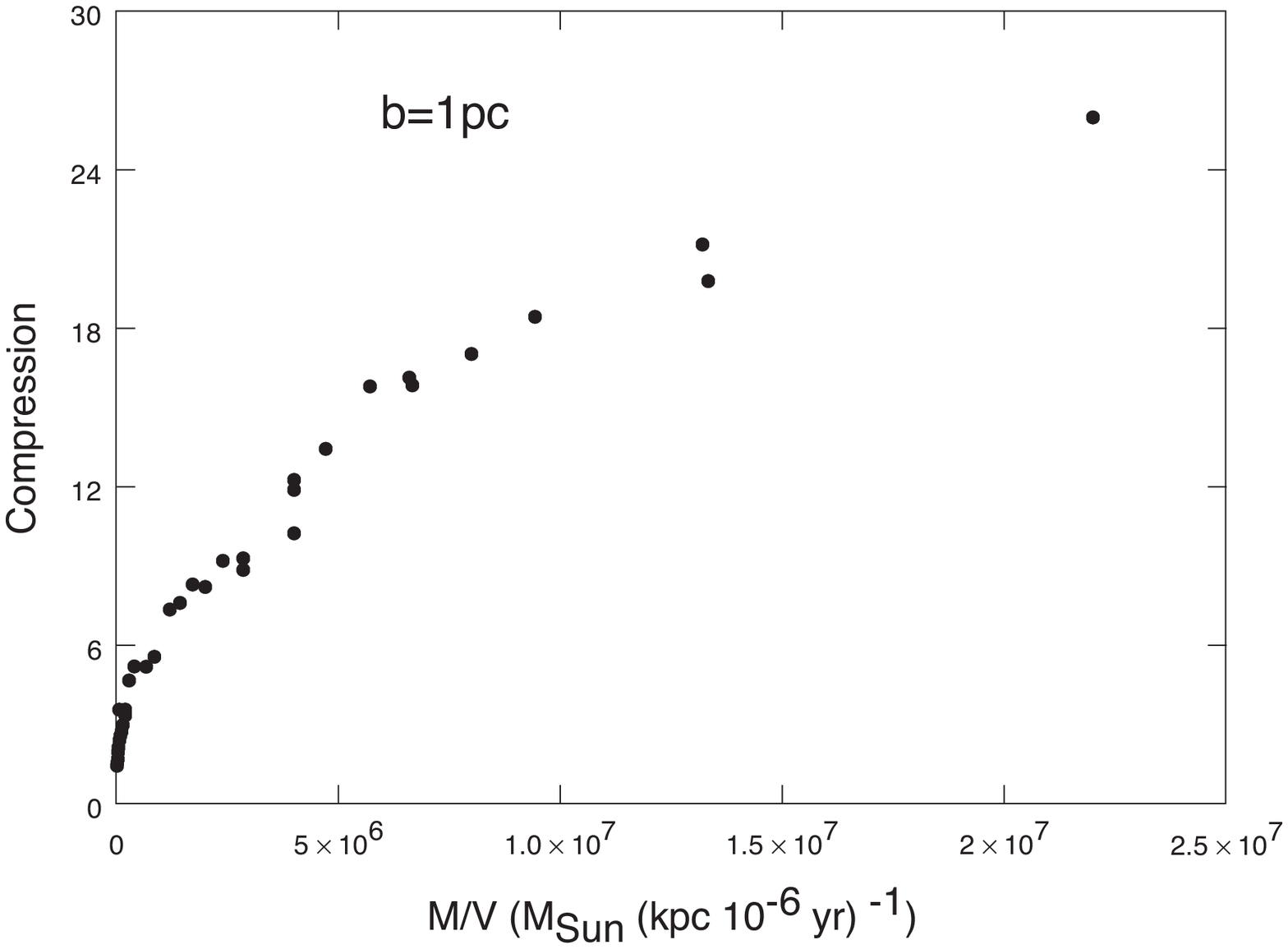}\\
\includegraphics[bb= 0 0 495 390,width=69mm]{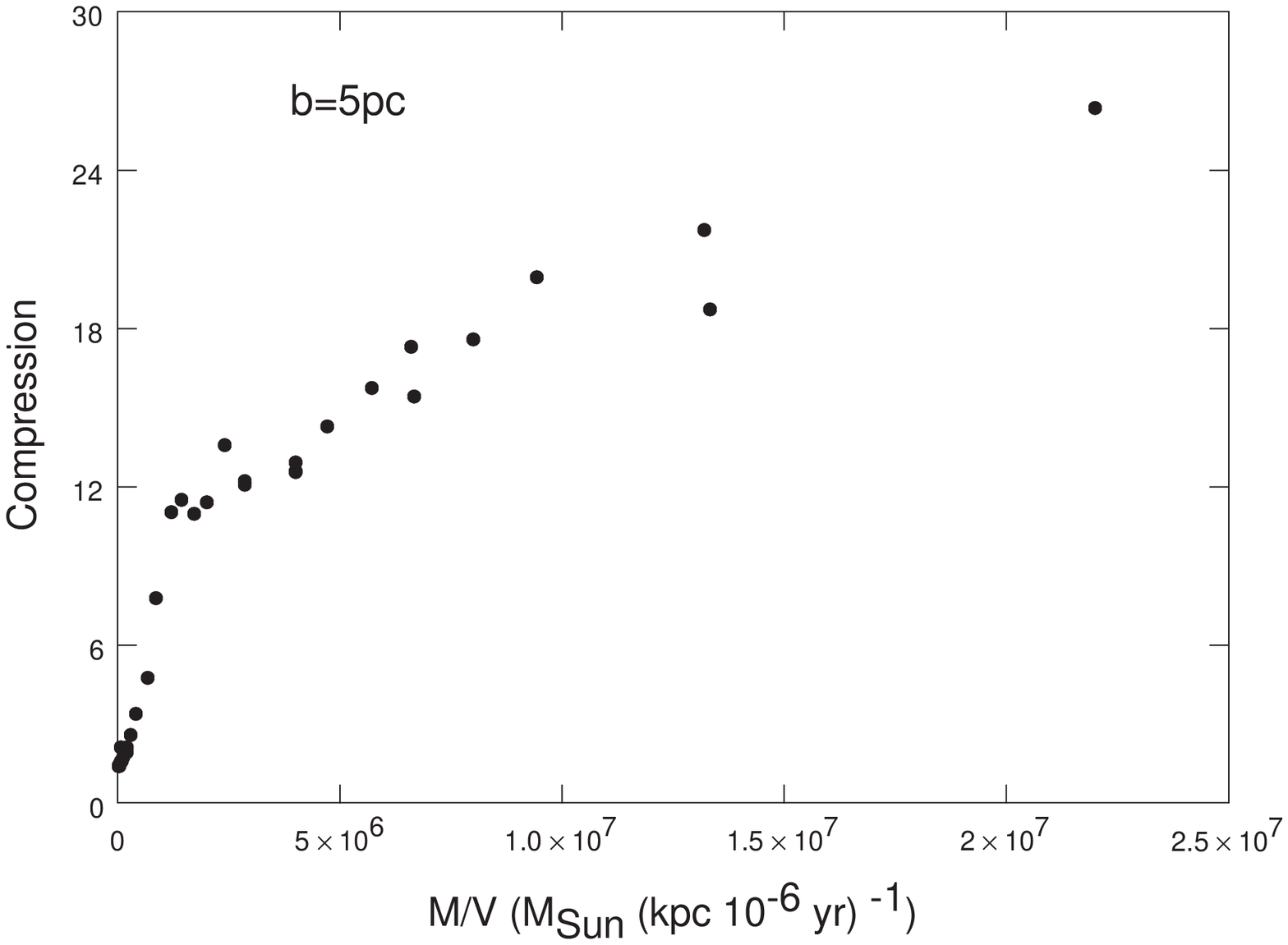}\\
\includegraphics[bb= 0 0 495 390,width=69mm]{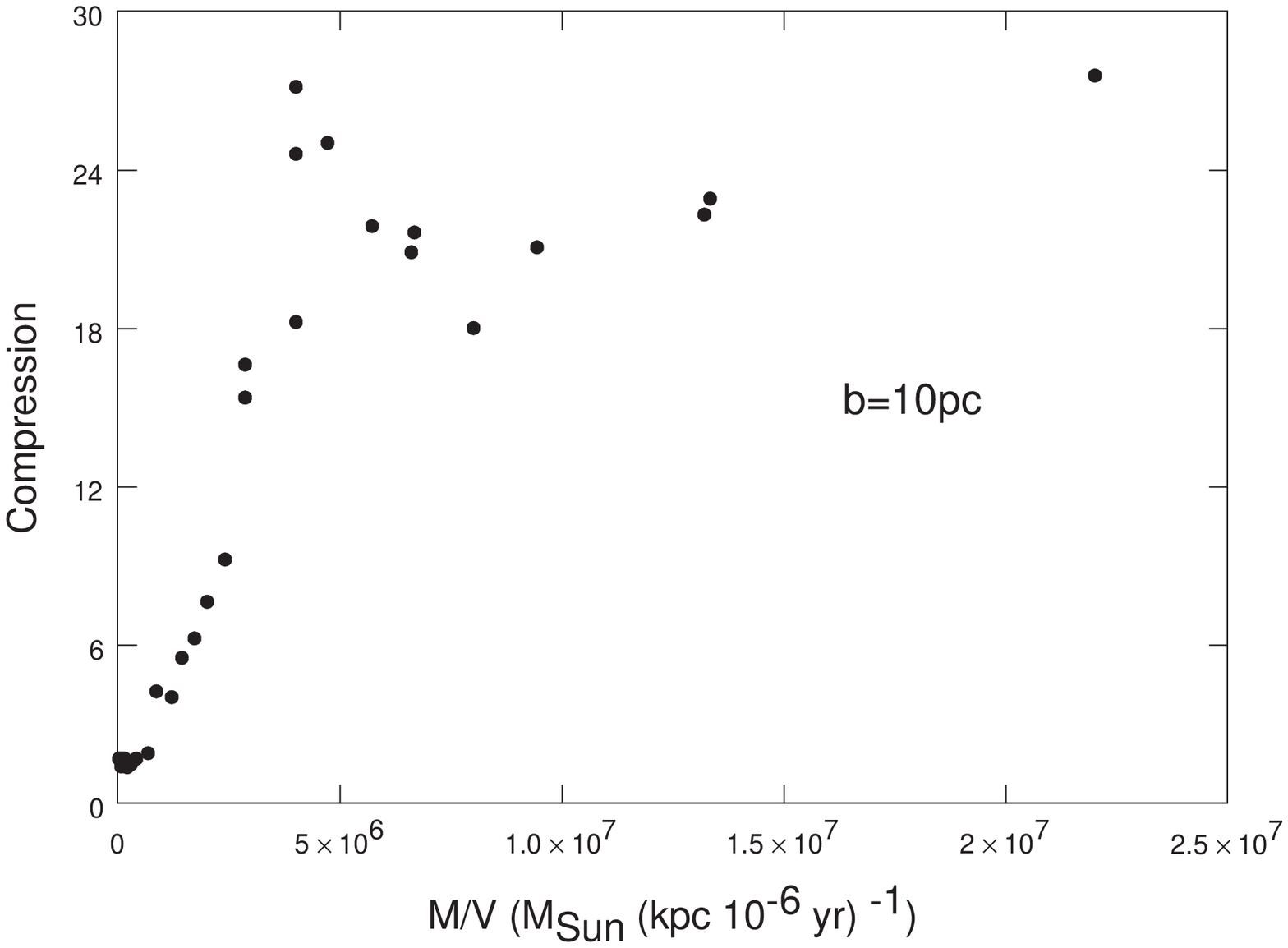}\\
\end{tabular}
%\end{center}
\caption{Compression as a function of GC mass divided by GC velocity, for three values of $b$}
\end{figure}

\subsection{Shock wave generation and material compression}

As modelled above, the test masses obey the collisionless Boltzmann equation, and we have shown that under such circumstances, gravitational focussing 
can occur.  Even if the GC disk crossing does cause 
gravitational focussing, the perturbation does 
not necessarily lead to star formation. This 
depends on the availability of molecular gas. 
Most of the Galaxy's molecular hydrogen is 
inhomogenously distributed and concentrated in a 
few thousand clouds of 10pc radius with masses 
$>$10$^{{\rm 5}}$ $M_{ \odot} $ 
(Binney \& Tremaine, 1994), though sizes reach up to 100pc with masses 
 $>$6 $\times$10$^{{\rm 6}}$ $M_{ \odot} $    
(Solomon et al., 1987). They are more prevalent in spiral 
arms, where they provide a fill factor of 
 0.1--0.5 (McKee \& Ostriker, 2007). 

In treating the impact on any gas in the disk, we need to include the 
additional collisional effects. The importance of this was 
first pointed out and investigated by Levy (2000).  Levy treats the GC with an associated Plummer potential, impinging vertically on a gas 
layer.  He uses a 2d hydrodynamical mesh code, and shows that a shock wave is set up by the GC.  After dissipation of the shock wave, 
there remains a disturbed cylindrical region of diameter up to 1kpc for a typical scale 
height of $ \sim 100$pc. Levy reports compression values in the shock wave 
of 2-30 for the different runs. If heating and cooling were allowed, these 
values could be larger.   Levy also finds that compression increases with increasing mass, with decreasing velocity and with smaller $b$.  We note that the compression values quoted here for the two mechanisms are broadly similar, with a range up to $\sim$30.  However, as the shock wave mechanism occurs in gas, it is expected to be more efficient at star formation.

\section{Locating impacts within the disk}
Since both the collisionless and collisional treatments, and particularly the latter when gas is present, tend to increase the concentration of disk material, we now examine whether there is any observational evidence of star formation as a result of such impacts.

\subsection{The sample of Globular Clusters}

The most widely used Galactic GC catalogue is that of Harris 
(1996), which is regularly updated: the latest version, 
available at www.physics.mcmaster.ca/Globular.html is dated 2003. This version of 
the Harris catalogue lists one hundred and fifty-five GCs. The catalogue does not contain proper motion information, and such data are difficult to determine reliably.   We therefore have narrowed our study to GCs with well-determined proper motion values from the subset of  Allen, 
Moreno \& Pichardo (2006 and 2008), except for NGC 6397, where we use more recent 
parameter values published since 2006 (Kalirai et al. 2007). The distances 
from the Sun, radial velocities, and proper motions for the fifty-four GCs 
are given in Table 3.

\newcommand{\PreserveBackslash}[1]{\let\temp=\\#1\let\\=\temp}
\let\PBS=\PreserveBackslash

\begin{table*}
\centering
\caption{Proper motions, radial velocities, heliocentric distances, and masses of selected GCs}
\begin{tabular}
{llr@{$\pm$}lr@{$\pm$}lr@{$\pm$}lr@{$\pm$}lr@{$\times$}lllllllll}
\hline
&Cluster& 
\multicolumn{2}{c}{ $\mu _{{\rm \alpha} }$ cos$\delta $}& 
\multicolumn{2}{c}{$\mu _{{\rm \delta} }$}& 
\multicolumn{2}{c}{ D to Sun}&
\multicolumn{2}{c}{$V_{{\rm r}{\rm a}{\rm d}}$}& 
\multicolumn{2}{c}{Mass}& 
\multicolumn{1}{c}{Refs \S  \par }& 
\\
& 
& 
\multicolumn{2}{c}{ mas y$^{{\rm -} {\rm 1}}$}& 
\multicolumn{2}{c}{mas y$^{{\rm -} {\rm 1}}$}& 
\multicolumn{2}{c}{ kpc}&
\multicolumn{2}{c}{km s$^{{\rm -} {\rm 1}}$}& 
\multicolumn{2}{c}{$M_{\odot}$}
\\

\hline
& 
NGC104& 
%$5.64\pm 0.2$& 
5.64& 0.2& 
-2.05& 0.2& 
 4.3& 0.35& 
-18.7& 0.2& 
1.0 &$10^{6}$& 
1, 2, 3, 12 \\
%\hline
& 
NGC288& 
4.4& 0.23& 
-5.62& 0.23& 
7.6& 0.35& 
-46.4& 0.4& 
8.5 &$10^{4}$& 
1, 12, 13 \\
%\hline
& 
NGC362& 
5.07& 0.714& 
-2.55& 0.72& 
7.8& 0.6& 
223.5& 0.5& 
3.9 &$ 10^{5}$& 
1, 12, 14 \\
%%\hline
& 
NGC1851& 
1.28& 0.68& 
2.39& 0.65& 
11.3& 1.3& 
320.5& 0.6& 
3.7 &$ 10^{5}$& 
1, 4, 5, 12 \\
%%\hline
& 
NGC1904& 
2.12& 0.64& 
-0.02& 0.64& 
\multicolumn{2}{l}{12.2\dag} & 
207.5& 0.5& 
2.4 & $ 10^{5}$& 
1, 12 \\
%%\hline
& 
NGC2298& 
4.05& 1& 
-1.72& 0.98& 
\multicolumn{2}{l}{\,\,\,9.3\dag} & 
149.4& 1.3& 
5.7 &$ 10^{4}$& 
1, 12 \\
%%\hline
& 
NGC2808& 
0.58& 0.45& 
2.06& 0.46& 
\multicolumn{2}{l}{\,\,\,9.6\dag} & 
93.6& 4& 
9.7 &$ 10^{5}$& 
21, 22, 27 \\
%%\hline
& 
NGC3201& 
5.28& 0.32& 
-0.98& 0.33& 
5.0& 0.4& 
494& 0.2& 
1.6 &$ 10^{5}$& 
21, 23, 27 \\
%%\hline
& 
NGC4147& 
-1.85& 0.82& 
-1.3& 0.82& 
16.4& 1.9& 
183& 1& 
5.0 &$ 10^{4}$& 
1, 6, 12 \\
%%\hline
& 
NGC4372& 
-6.49& 0.33& 
3.71& 0.32& 
5.8& 0.5& 
72.3& 1.2& 
2.2 &$ 10^{5}$& 
21, 24, 25, 27 \\
%%\hline
& 
NGC4590& 
-3.76&0.66& 
1.79&0.62& 
\multicolumn{2}{l}{\,\,\,8.9\dag} & 
-95.1& 0.6& 
1.5 &$ 10^{5}$& 
1, 12 \\
%%\hline
& 
NGC4833& 
-8.11& 0.35& 
-0.96& 0.34& 
\multicolumn{2}{l}{\,\,\,6.5\dag} & 
200.2& 1.2& 
3.1 &$ 10^{5}$& 
21, 25, 27 \\
%%\hline
& 
NGC5024& 
0.5& 1& 
-0.1& 1& 
\multicolumn{2}{l}{17.9\dag} & 
-79.1& 4.1& 
5.2 &$ 10^{5}$& 
1, 12 \\
%%\hline
& 
NGC5139& 
-5.08& 0.35& 
-3.57& 0.34& 
4.9& 0.3& 
232.5& 0.4& 
3.3 &$ 10^{6}$& 
1, 12, 15 \\
%%\hline
& 
NGC5272& 
-1.1& 0.51& 
-2.3& 0.54& 
9.5& 1& 
-147.1& 0.4& 
6.4 &$ 10^{5}$& 
1, 6, 12 \\
%%\hline
& 
NGC5466& 
-4.65& 0.82& 
0.8& 0.82& 
15.4& 1.6& 
107.7& 0.3& 
1.0 &$ 10^{5}$& 
1, 6, 12 \\
%%\hline
& 
NGC5897& 
-4.93& 0.86& 
-2.33& 0.84& 
12.4& 1.7& 
101.7& 1& 
1.3 &$ 10^{5}$& 
1, 12, 16 \\
%%\hline
& 
NGC5904& 
5.07& 0.68& 
-10.7& 0.56& 
7.2& 0.8& 
51.8& 0.5& 
5.7&$ 10^{5}$& 
1, 6, 12 \\
%%\hline

& 
NGC5927& 
-5.72& 0.39& 
-2.61& 0.40& 
\multicolumn{2}{l}{\,\,\,7.6\dag} & 
-107.5& 1.0& 
2.2 &$ 10^{5}$& 
21, 26, 27 \\
%%\hline

& 
NGC5986& 
-3.81& 0.45& 
-2.99& 0.37& 
10.4& 0.5& 
88.9& 3.7& 
4.1&$ 10^{5}$& 
21, 26, 27, 28 \\
%%\hline
& 
NGC6093& 
-3.31& 0.58& 
-7.2& 0.67& 
%8.3& & 
\multicolumn{2}{l}{\,\,\,8.3\dag} &
7.3& 4.1& 
3.3 &$ 10^{5}$& 
1, 12 \\
%%\hline
& 
NGC6121& 
-12.5& 0.36& 
-19.93& 0.49& 
\multicolumn{2}{l}{\,\,\,1.8\dag} & 
70.4& 0.4& 
1.3 & $10^{5}$& 
1, 12 \\
%%\hline
& 
NGC6144& 
-3.06& 0.64& 
-5.11& 0.72& 
\multicolumn{2}{l}{\,\,\,\,\,\,\,9\dag} & 
189.4& 1.1& 
8.6 &$ 10^{4}$& 
1, 12 \\
%%\hline
& 
NGC6171& 
-0.7& 0.9& 
-3.1& 1& 
\multicolumn{2}{l}{\,\,\,5.9\dag} & 
-33.8& 0.3& 
1.2 &$ 10^{5}$& 
1, 12 \\
%%\hline
& 
NGC6205& 
-0.9& 0.71& 
5.5& 1.12& 
6.8& 0.7& 
-246.6& 0.9& 
5.2 &$ 10^{5}$& 
1, 6, 12 \\
%%\hline
& 
NGC6218& 
1.3& 0.58& 
-7.83& 0.62& 
\multicolumn{2}{l}{\,\,\,\,4.2\dag} & 
43.5& 0.6& 
1.4 &$ 10^{5}$& 
1, 12 \\
%%\hline
& 
NGC6254& 
-6& 1& 
-3.3& 1& 
\multicolumn{2}{l}{\,\,\,\,4.1\dag} & 
75.5& 1.1& 
1.7 &$ 10^{5}$& 
1, 12 \\
%%\hline
& 
NGC6266& 
-3.5& 0.37& 
-0.82& 0.37& 
6.9& 0.7& 
-70& 1.3& 
8.1&$ 10^{5}$& 
12, 17 \\
%%\hline
& 
NGC6304& 
-2.59& 0.29& 
-1.56& 0.29& 
6.1& 0.6& 
-107.3& 3.6& 
1.5 &$ 10^{5}$& 
12, 17 \\
%%\hline
& 
NGC6316& 
-2.42& 0.63& 
-1.71& 0.56& 
11& 1.1& 
71.5& 8.9& 
3.7 &$ 10^{5}$& 
12, 17 \\
%\hline
& 
NGC6341& 
-3.3& 0.55& 
-0.33& 0.7& 
\multicolumn{2}{l}{\,\,\,\,7.4\dag} & 
-120.5& 1.7& 
3.3 &$ 10^{5}$& 
1, 12 \\
%%\hline
& 
NGC6362& 
-3.09& 0.46& 
-3.83& 0.46& 
\multicolumn{2}{l}{\,\,\,\,6.8\dag }& 
-13.3& 0.6& 
1.0 &$ 10^{5}$& 
1, 12 \\
%%\hline
& 
NGC6397& 
3.56& 0.04& 
-17.34& 0.04& 
2.6& 0.13& 
18.36& 0.13& 
7.7 &$ 10^{4}$& 
12, 18 \\
%%\hline
& 
NGC6522& 
6.08& 0.2& 
-1.83& 0.2& 
7.8& 0.8& 
-21.1& 3.4& 
2.0 &$ 10^{5}$& 
12, 17 \\
%%\hline
& 
NGC6528& 
-0.35& 0.23& 
0.27& 0.26& 
9.1& 0.9& 
184.9& 3.8& 
7.2 &$ 10^{4}$& 
12, 17 \\
%%\hline
& 
NGC6553& 
2.5& 0.07& 
5.35& 0.08& 
5.6& 0.6& 
-6.5& 2.7& 
2.2 &$ 10^{5}$& 
12, 17 \\
%\hline
& 
NGC6584& 
-0.22& 0.62& 
-5.79& 0.67& 
\multicolumn{2}{l}{\,12.9\dag} & 
222.9& 15& 
2.0 &$ 10^{5}$& 
1, 12 \\
%\hline
& 
NGC6626& 
0.3& 0.5& 
-3.4& 0.9& 
\multicolumn{2}{l}{\,\,\,\,5.2\dag} & 
15.8 & 1& 
3.2 &$ 10^{5}$& 
1, 12 \\
%\hline
& 
NGC6656& 
8.6& 1.3& 
-5.1& 1.3& 
\multicolumn{2}{l}{\,\,\,\,3.1\dag }& 
-149.1& 0.6& 
4.3 &$ 10^{5}$& 
1, 12 \\
%%\hline
& 
NGC6712& 
4.2& 0.4& 
-2& 0.4& 
\multicolumn{2}{l}{\,\,\,\,6.5\dag }& 
-107.7& 0.6& 
1.7 &$ 10^{5}$& 
1, 12 \\
%\hline
& 
NGC6723& 
-0.17& 0.45& 
-2.16& 0.5& 
8.8& 0.9& 
-94.5& 3.6& 
2.3 &$ 10^{5}$& 
12, 17 \\
%%\hline
& 
NGC6752& 
-0.69&0.42& 
-2.85& 0.45& 
3.8& 0.19& 
-32.1& 1.5& 
2.1 &$ 10^{5}$& 
1, 12, 19 \\
%%\hline
& 
NGC6779& 
0.3&1& 
1.4& 0.1& 
9.4& 1& 
-135.9& 0.9& 
1.5 &$10^{5}$& 
1, 6, 12 \\
%%\hline
& 
NGC6809& 
-1.42& 0.62& 
-10.25& 0.64& 
\multicolumn{2}{l}{\,\,\,\,\,\,\,\,5\dag} & 
174.9& 0.4& 
1.8 &$ 10^{5}$& 
1, 12 \\
%%\hline
& 
NGC6838& 
-2.3& 0.8& 
-5.1& 0.8& 
3.6& 0.3& 
-22.9& 0.2& 
3.0 &$ 10^{4}$& 
1, 12, 20 \\
%%\hline
& 
NGC6934& 
1.2& 1& 
-5.1& 1& 
14.9& 1.5& 
-412.2& 1.6& 
1.6 &$ 10^{5}$& 
1, 6, 12 \\
%%\hline
& 
NGC7006& 
-0.96&0.35& 
-1.14& 0.4& 
40& 2& 
-384& 0.4& 
2.0 &$ 10^{5}$& 
8, 12 \\
%%\hline
& 
NGC7078& 
-0.95& 0.51& 
-5.63& 0.5& 
9.5& 1.1& 
-106.6& 0.6& 
8.0 &$ 10^{5}$& 
1, 6, 12 \\
%%\hline
& 
NGC7089& 
5.9& 0.86& 
-4.95& 0.86& 
11.2& 1.2& 
-3.1& 0.9& 
6.9 &$ 10^{5}$& 
1, 6, 12 \\
%%\hline
& 
NGC7099& 
1.42& 0.69& 
-7.71& 0.65& 
\multicolumn{2}{l}{\,\,\,\,7.3\dag} & 
-184.3& 1& 
1.6 &$ 10^{5}$& 
1, 12 \\
%%\hline
& 
PAL3& 
0.33& 0.23& 
0.3& 0.31& 
\multicolumn{2}{l}{81.6\dag} & 
83.4& 8.4& 
3.3 &$ 10^{5}$& 
1 \\
%%\hline
& 
PAL5& 
-1.78& 0.17& 
-2.32& 0.22& 
9.9& 2.2& 
-55& 16& 
2.0 &$ 10^{4}$& 
1, 7, 12 \\
%%\hline
& 
PAL12& 
-1.2& 0.3& 
-4.21& 0.29& 
19.5& 0.9& 
27.8& 1.5& 
1.1 &$ 10^{4}$& 
9, 12 \\
%%\hline
& 
PAL13& 
2.3& 0.26& 
0.27& 0.25& 
24.3& 1.1& 
-28& 1.2& 
5.4 &$ 10^{3}$& 
10, 11, 12 \\
\hline
\end{tabular}
\\  \S Refs: 1, Dinescu et al., 1999; 2, McLaughlin et al., 2006; 3, 
Anderson \& King, 2004; 4, Stetson, 1981; 5, Walker, 1998; 6, Dauphole \& 
Colin, 1995; 7, Scholz et al., 1998; 8, Dinescu et al, 2001; 9, Dinescu et 
al., 2000; 10, Siegel et al., 2001; 11, C\^{o}t\'{e} et al, 2002; 12 Allen, 
Moreno, and Pichardo, 2006; 13, Alcaino et al, 1997; 14, Sz\'{e}kely et al., 
2007; 15, van de Ven et al, 2006; 16, Sarajedini , 1992; 17 Dinescu et al., 
2003; 18 Kalirai et al., 2007; 19, Renzini et al, 1996; 20, Grundahl et al, 
2002; 21, Allen et al., 2008; 22, Faulkner et al 1991; 23, C\^{o}t\'{e} et al., 1995; 24, Hartwick \& Hesser, 1973; 25, Geisler et al., 1995; 26, Harris 1996; 27 Casetti-Dinescu et al., 2007; 28, Alves et al., 2001\\ $\dag$ No uncertainties on distance could be found for these clusters.
\\

\end{table*}
\begin{figure*}
\begin{center}
\begin{tabular}{c@{\hspace{1pc}}c}
\includegraphics[bb=4 4 476 460,height=8.3cm,clip]{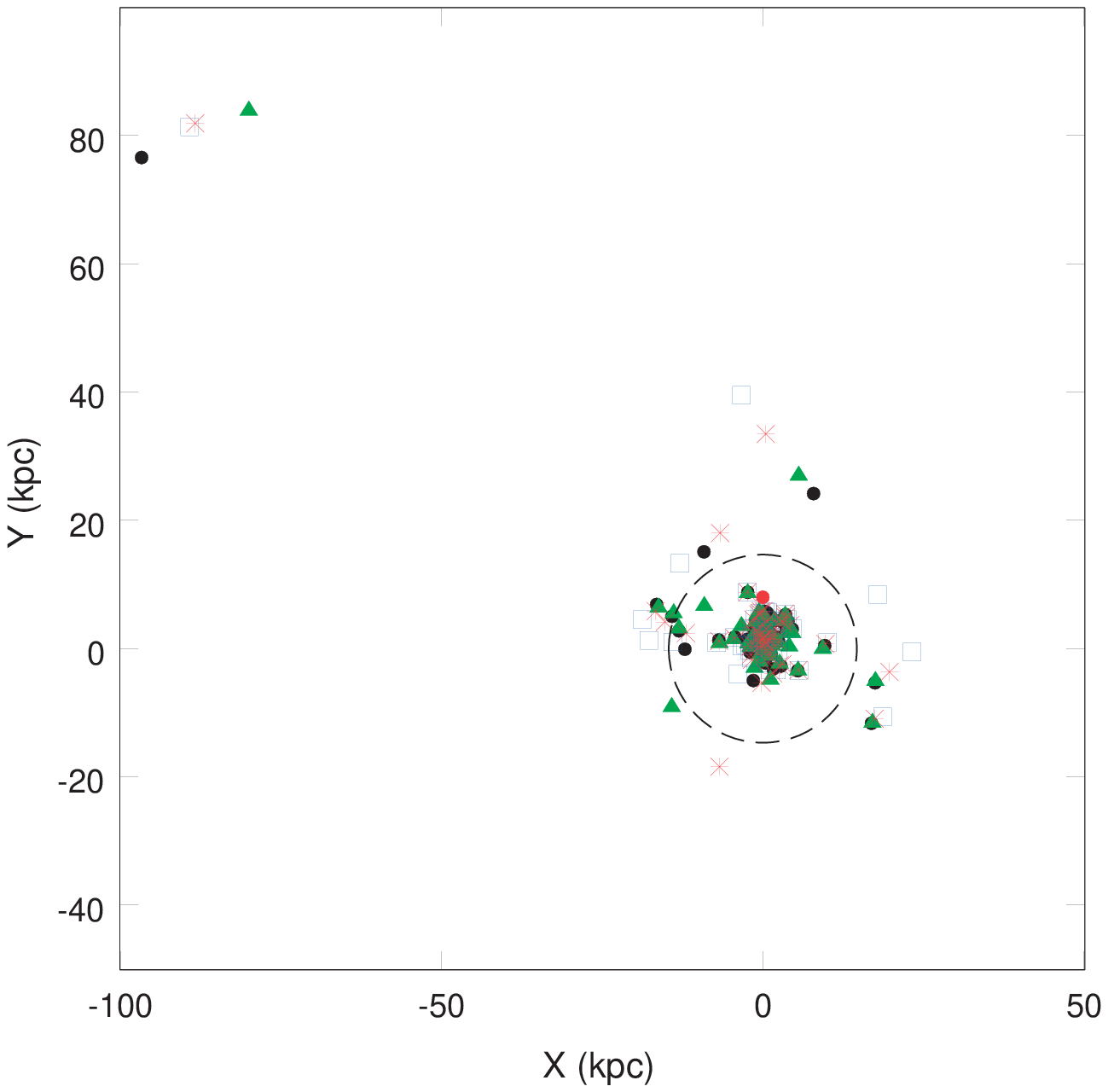} &
\includegraphics[bb=4 4 476 469,height=8.3cm,clip]{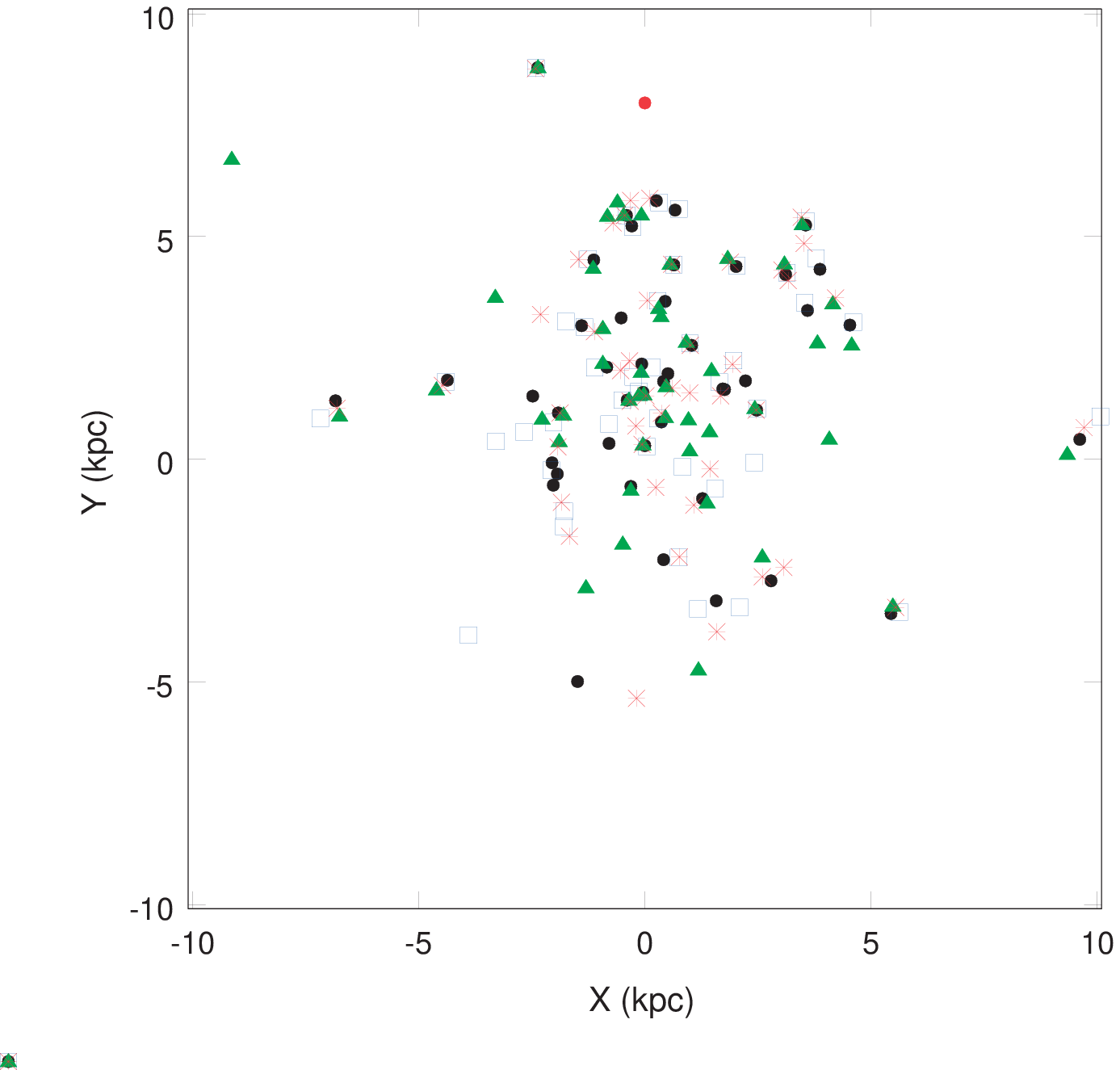} \\
\end{tabular}
\end{center}
\caption{Present-day positions of latest crossing for the four 
potentials and fifty-four GCs in the Galactic plane. The Sun (red circle) 
is at $x= 0$ kpc, $y= 8$ kpc. Potential: DA, filled black circles; FE, red stars; FL, 
green triangles; PA, blue open squares.  The black circle of 15 kpc radius centered on the Galactic centre represents the extent of the visible disk (Sparke \& Gallagher, 2000).  The plot on the right shows an expanded region in the disk.}
\end{figure*}

\subsection{The location of impacts}

We used the four potentials and all fifty-four GC initial data to calculate 
their prior orbits. From this we have determined the location 
and time of the most recent crossing of the Galactic plane. Then we 
used the circular velocity at the GC Galactic radius for the potential in use to 
determine the present-day location of the impact points, knowing the time 
since the crossing. The resulting present-day positions of these crossings 
are shown in Figure 4. A few cross outside the visible Galactic disk, but most pass through it.  Note that the four points that cluster around $x=-90$ kpc, $y=80$ kpc in the left-hand panel correspond to Palomar 3.  The right-hand panel concentrates on the Galactic disk.  The average GC speed at time of crossing is $\sim$280 kms$^{{\rm-}{\rm 1}}$ with a standard deviation of $\sim$90 kms$^{{\rm-}{\rm 1}}$.

\begin{table*}
\centering
\caption{Present day distance D of last crossing point from Sun (kpc) and times T since latest 
crossing (Myr), for the four potentials, for the nominal kinematic parameters in Table 3.  Uncertainties in position of impact point are provided in Figure 7, and discussed in Section 3.2.2.}
\begin{tabular}
{lr@{.}lr@{.}lr@{.}lr@{.}lr@{.}lr@{.}lr@{.}lr@{.}lr@{.}l}

\hline
& 
\multicolumn{4}{c}{DA} & 
\multicolumn{4}{c}{FE} & 
\multicolumn{4}{c}{FL} & 
\multicolumn{4}{c}{PA}  
\\
&\multicolumn{2}{c}{ D (kpc)}& 
\multicolumn{2}{c}{ T (Myr)}& 
\multicolumn{2}{c}{D (kpc)}& 
\multicolumn{2}{c}{T (Myr)}& 
\multicolumn{2}{c}{D (kpc)}& 
\multicolumn{2}{c}{T (Myr)}& 
\multicolumn{2}{c}{D (kpc)}& 
\multicolumn{2}{c}{T (Myr)} 
 \\
\hline
NGC104& 
 2&2& 
48&$8\pm2.1$&
 2&2& 
46&$7\pm1.9$& 
 2&3& 
52&$6\pm1.9$& 
 2&3& 
49&$1\pm2.3$ \\

NGC1851& 
2&5& 
43&$2\pm8.0$& 
2&1& 
42&$5\pm7.8$& 
2&5& 
43&$6\pm8.1$& 
2&5& 
44&$3\pm8.7$ \\

NGC3201& 
2&5& 
5&$5\pm0.4$& 
2&5& 
5&$4\pm0.4$& 
2&5& 
5&$5\pm0.4$& 
2&5& 
5&$5\pm0.4$ \\

NGC6397& 
2&6& 
3&$7\pm0.2$& 
2&6& 
3&$7\pm0.2$& 
2&6& 
3&$7\pm0.2$& 
2&6& 
3&$7\pm0.2$ \\

NGC6752& 
2&8& 
29&$4\pm1.7$& 
2&8& 
29&$5\pm1.6$& 
2&7& 
35&$3\pm2.2$& 
2&8& 
29&$3\pm1.8$ \\

NGC6838& 
4&5& 
15&$6\pm0.5$& 
4&3& 
14&$6\pm0.5$& 
4&4& 
18&$6\pm0.7$& 
4&4& 
14&$7\pm0.5$ \\

\hline
\end{tabular}
\end{table*}

\subsubsection{Effect of the potential model}

Systematic errors in impact location arise from our uncertain knowledge of the Galactic parameters, 
such as the extent and mass of components making up the potential, and from more detailed configuration effects, such 
as the bar and spiral structure that are not reflected in the  homogeneous bulges and disks. We neglect these latter effects as there are still uncertainties 
about the structure of the bar and arms.  However we can 
gauge the effect of the former uncertainties in extent and mass of components by 
comparing all four axisymmetric Galactic potentials. For example, in Figure 4 (left-hand panel), 
the four points at $x\sim0$, $y>$20 kpc correspond to the crossing point of NGC 5024, as 
calculated with the four potentials. There is a significant scatter of $\sim$ 20kpc in the impact position.  The crossing occurred $\sim$200Myr ago. 
Figure 5 shows a restricted part of the Galactic plane in the region of the Sun. This illustrates 
that the more concentrated impact points correspond to the more recent crossings, 
NGC 104 at $\sim $47Myr, NGC 1851 at $\sim $43Myr, NGC 3201 at $\sim $5.4Myr, NGC 6397 at $\sim $3.7Myr, 
NGC 6752 at $\sim $30Myr, and NGC 6838 at $\sim $15Myr.

We will concentrate our discussion on the six cases noted in Figure 5. 
They are selected because NGC 104/ 1851/ 3201/ 6397/ 6752 impact points are the 
closest to us, and thus afford the best visibility. We add the impact point location of NGC 6838, as the path between it and us crosses a less dense region of the 
disk. In all six cases, the times since impact are short enough for massive 
stars still to be seen.  Table 4 provides the times of crossing 
and the relevant distances from the Sun for these six GCs for the nominal values of initial conditions in Table 3.  The uncertainties in the position of the impact points are discussed in Section 3.2.2.  The corresponding uncertainties in time since impact are shown in Table 4, and their origin are also discussed in Section 3.2.2.

\subsubsection{Effect of uncertainties in measured parameters}

Uncertainties (random errors) in the values of the Sun's motion in the LSR, 
velocity of the LSR, distances, radial velocities, and proper motions 
affect the error in disk crossing location. We have therefore made a calculation that includes the effect of all these uncertainties.  To appreciate the effect of these 
uncertainties on the impact sites of the six selected GCs, we assume that the 
values for Solar motion, LSR motion, distances, radial velocities and proper 
motions are all normally distributed, with standard deviations equal to the 
quoted experimental errors. We then performed a Monte Carlo simulation for each of the six selected GCs, and 20,000 trials per GC.  Figure 6 shows the combined effect of the parameter uncertainties and different potentials, at the 90\% level.  This also shows that the effect of the potential differences can be as large as the effect of the orbital uncertainties, specifically in the case of NGC104.  The combined results for the FE potential appear in Figure 7, as 90$\%$ and 68$\%$ probability contour levels.    We note that, as expected, the values of uncertainties in time since impact, and the size of the area encompassed by the 90\% contours rank in the same order.

\begin{figure}
\includegraphics[bb=23 40 580 440, width=99mm,clip]{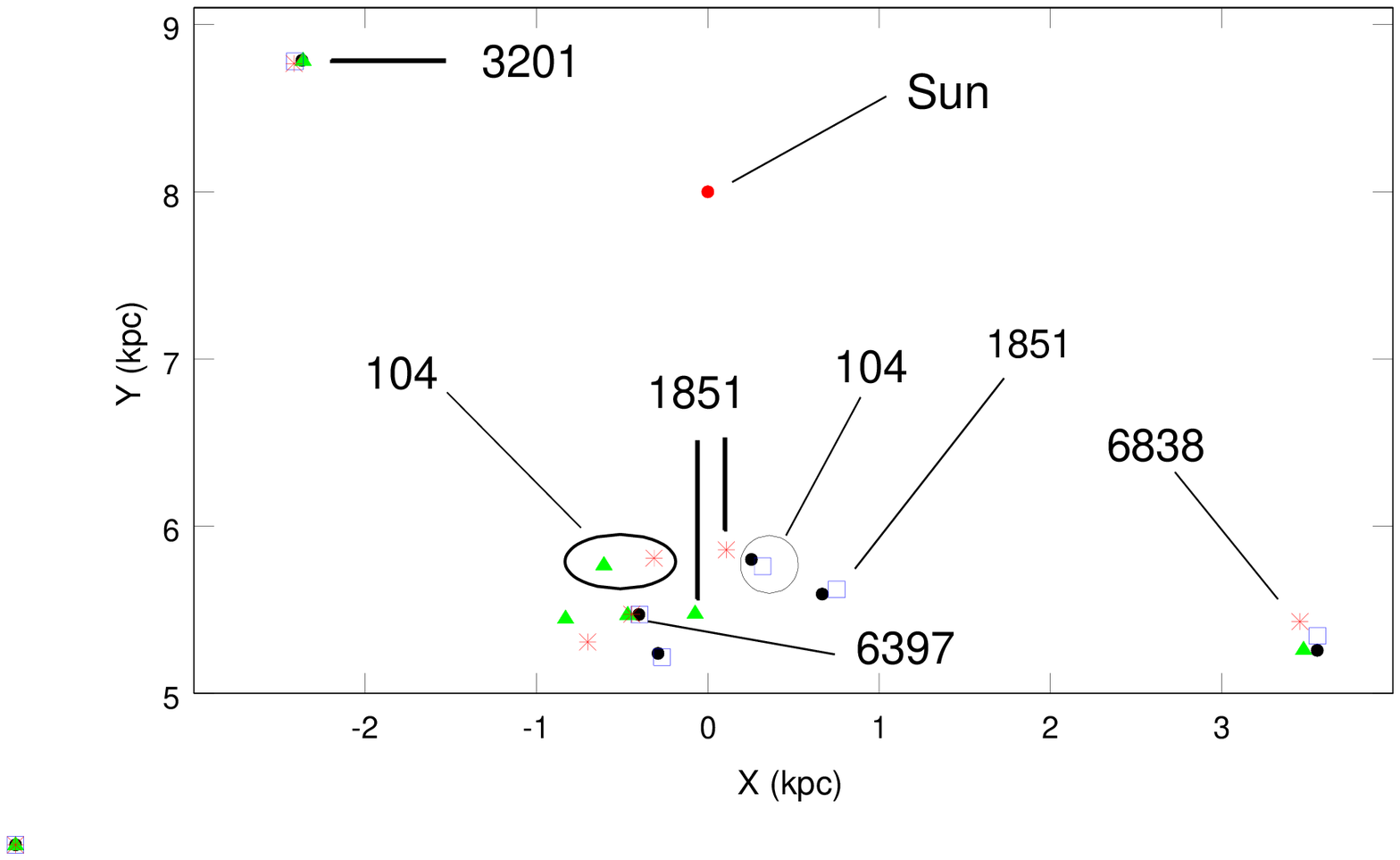}
\caption{Restricted portion of Figure 4; unmarked positions relate 
to NGC 6752}
\end{figure}

\section{Search areas for remnants of impacts}

Here we explore the six search areas identified above.

\subsection{Is NGC 6231 a remnant of an NGC6397 impact?}

The only claim to have discovered an impact remnant  -- a 
cluster -- in the disk relates to one of the six cases examined here: NGC 6397. 
Rees \& Cudworth, cited in Wright (2004), suggested the site of the last 
impact of NGC 6397 to correspond to the open cluster NGC 6231 ($\textit{l}$ = 
$343.46 \deg$, $b$ = +1.186 $\deg$). The cluster has an angular diameter of 15\'{} (Piatti, Clari\'{a} \& Bica, 1998).  We calculate its impact time to be $\sim $3.7Myr ago, consistent with Rees and Cudworth's estimate of less than 5Myr. We can compare our time since last crossing at 3.7Myr with further estimates of the age of NGC 3201 in the literature.  Piatti, Clari\'{a} \& Bica (1998) estimate it to be 5Myr.   van den Ancker  (2007)  estimates the age to be $\sim$3Myr.  Sana et al. (2007) derive an age of at least 10Myr, based on examining the low-mass stars in the cluster, but they also indicate that low-mass star formation was followed by a starburst 1--4 Myr ago.  (Interestingly, van den Ancker (2007)  concludes that the stellar population in NGC6231 is not incompatible with star formation being induced by GC impact on the disk, given that the cluster appears to have few low-mass stars, which he suggests may be the signature of a violent origin).  Thus apart from Sana's determination, these ages are consistent with the last crossing time. However, we do not confirm the link established by Rees \& Cudworth.  Figure 8 shows the position of NGC 6231 and the 90$\%$ level for the NGC 6397 impacts, at the time of impact 3.7 Myr ago. Also shown is the contemporaneous position of NGC 6231 (black square) and the points corresponding to that position $\pm$3$\sigma$ from that position, based on Sana et al.'s (2007) distance estimate of 1637$\pm$30 pc.  Note that NGC 6231 is $\sim$7pc in diameter.  Clearly, NGC 6231 is not a remnant of the NGC 6397 impact, being $\sim$0.6kpc away from the contour.  Regrettably, the 
proper motions used by Rees \& Cudworth were never published, preventing any 
further comparison.

\begin{figure*}
\begin{center}
\begin{tabular}{c@{\hspace{1pc}}c}
\includegraphics[bb=30 42 490 469,height=7.5cm,clip=]{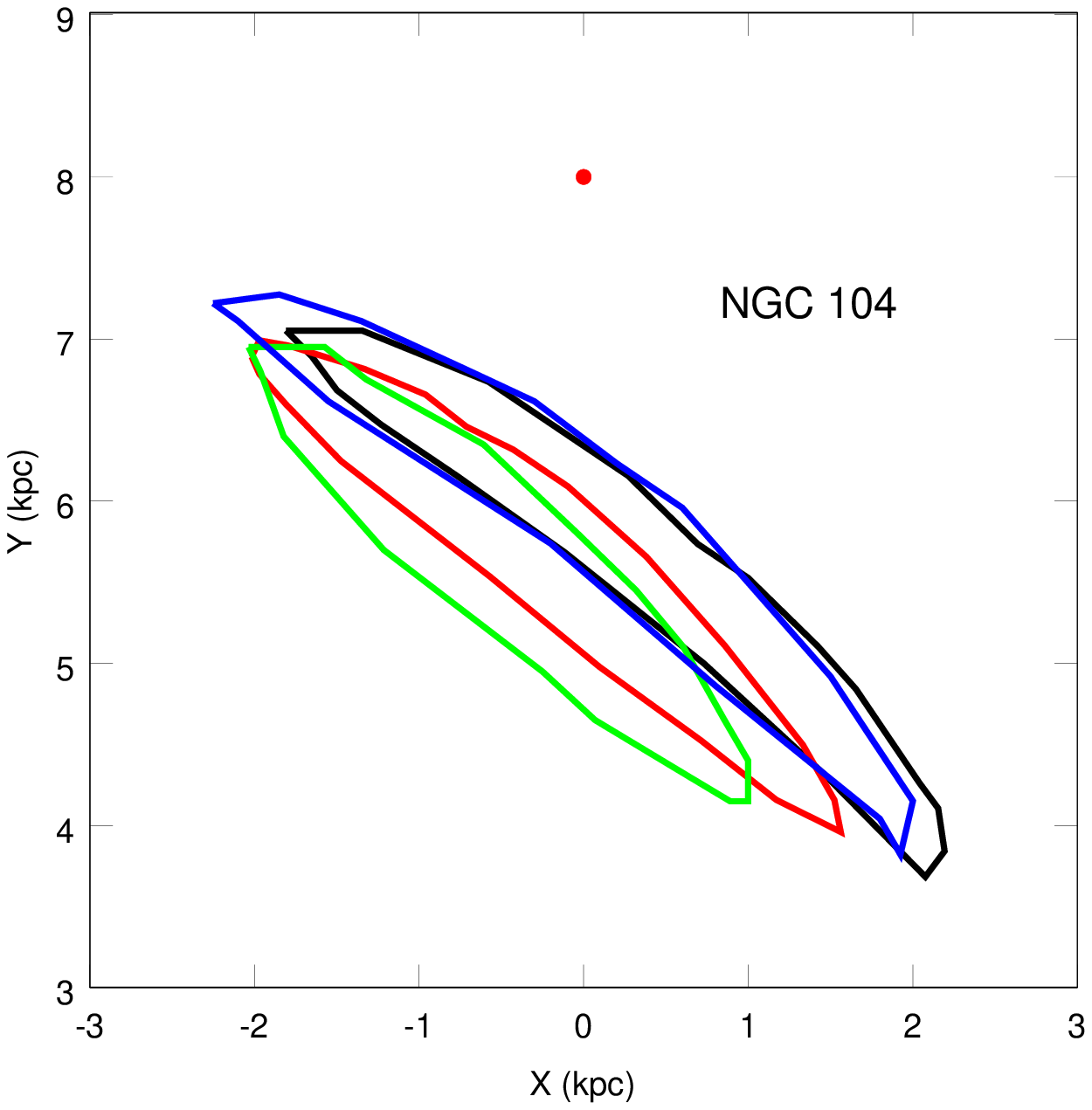} &
\includegraphics[bb=30 42 490 469,height=7.5cm,clip=]{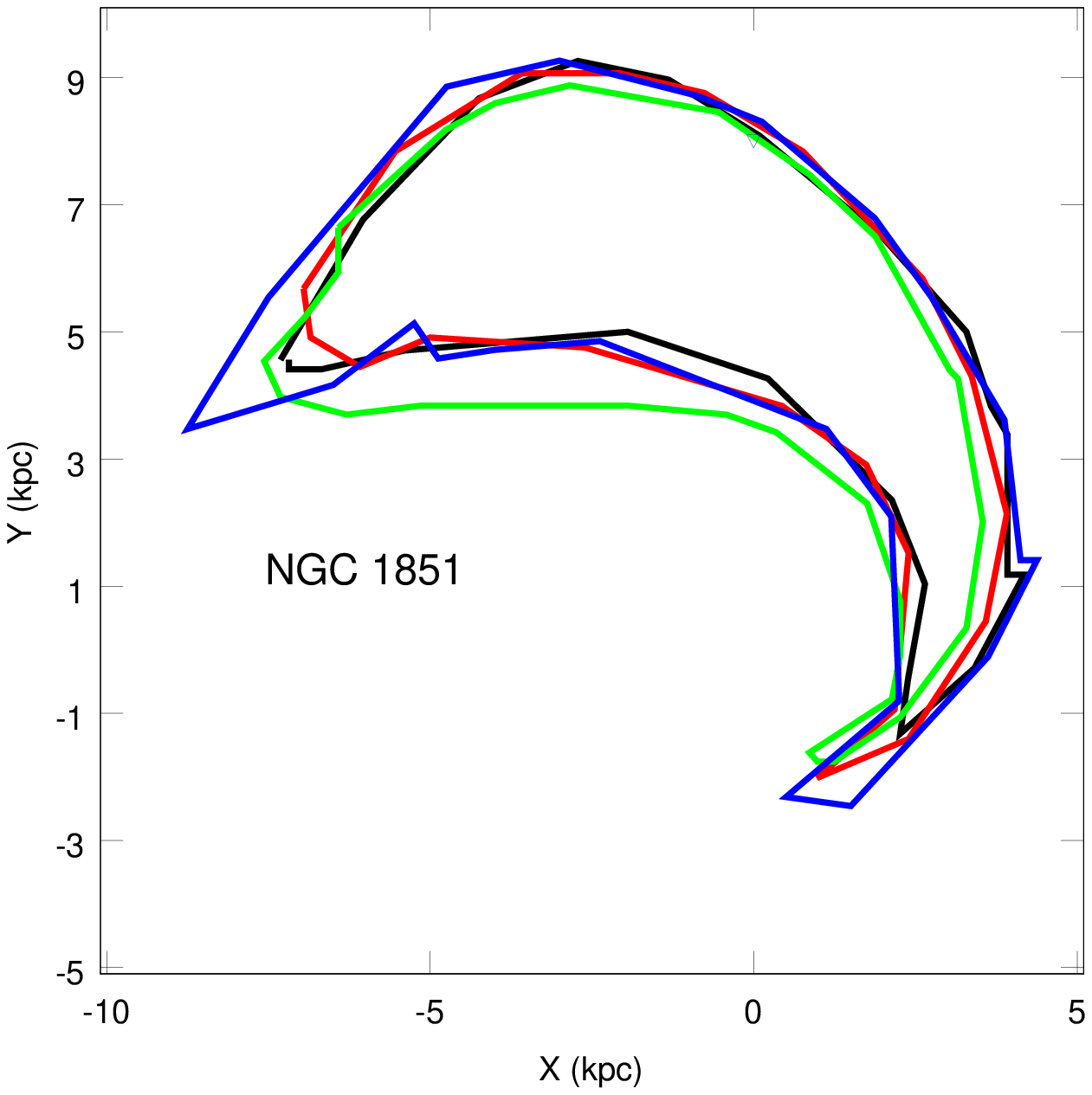} \\
\includegraphics[bb=30 42 490 469,height=7.5cm,clip=]{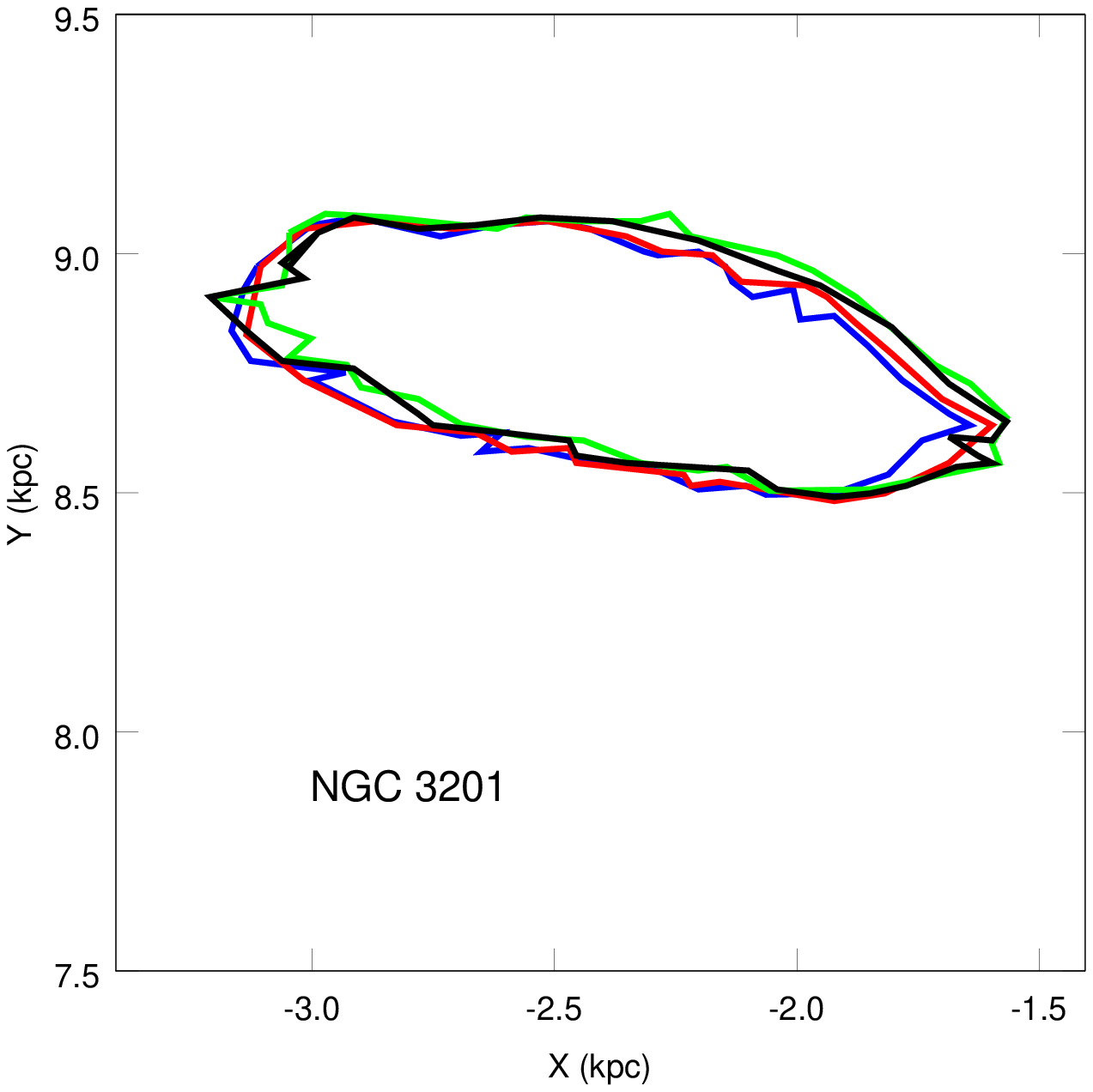} &
\includegraphics[bb=30 42 490 469,height=7.5cm,clip=]{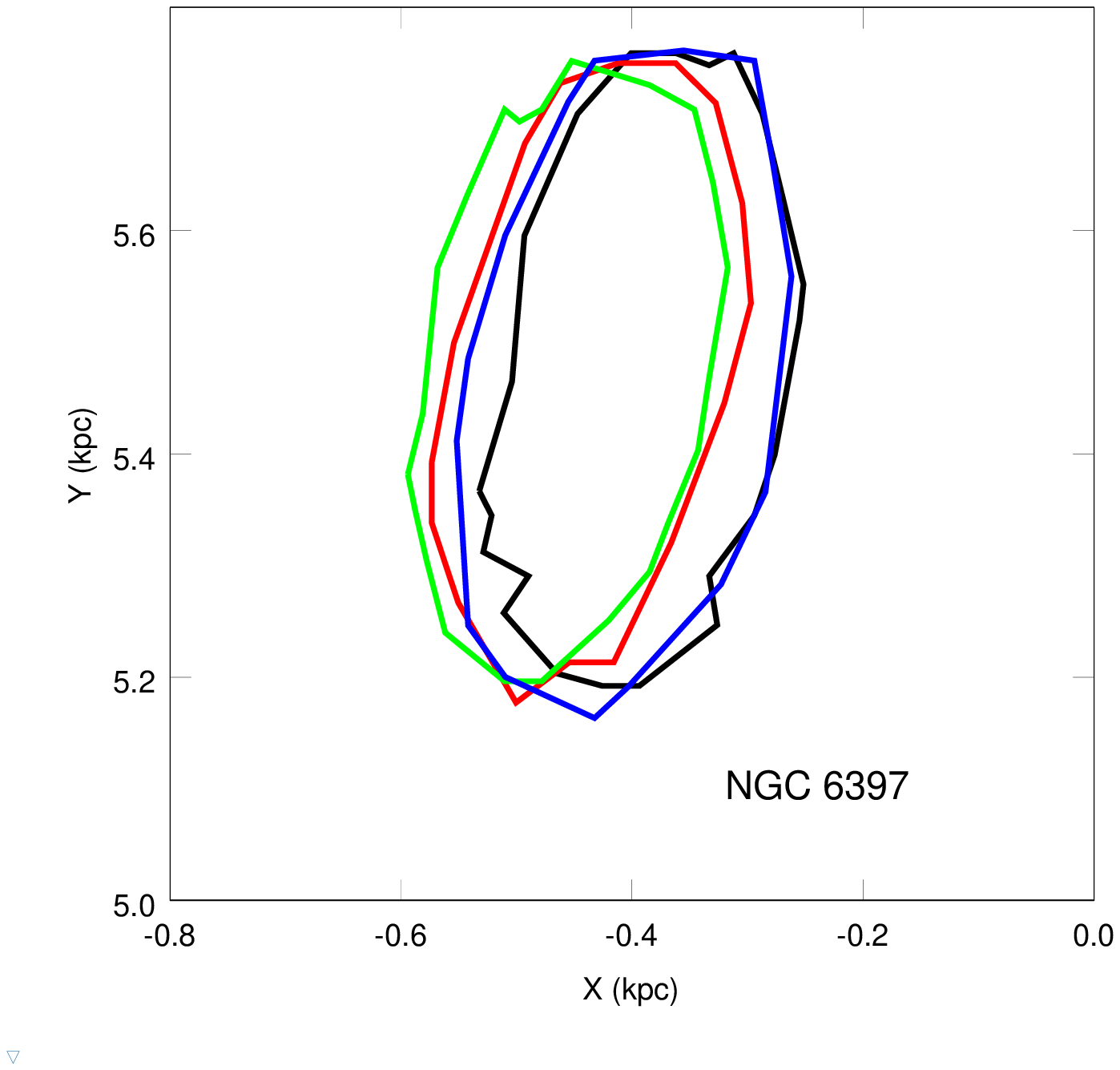} \\
\includegraphics[bb=30 42 490 469,height=7.5cm,clip=]{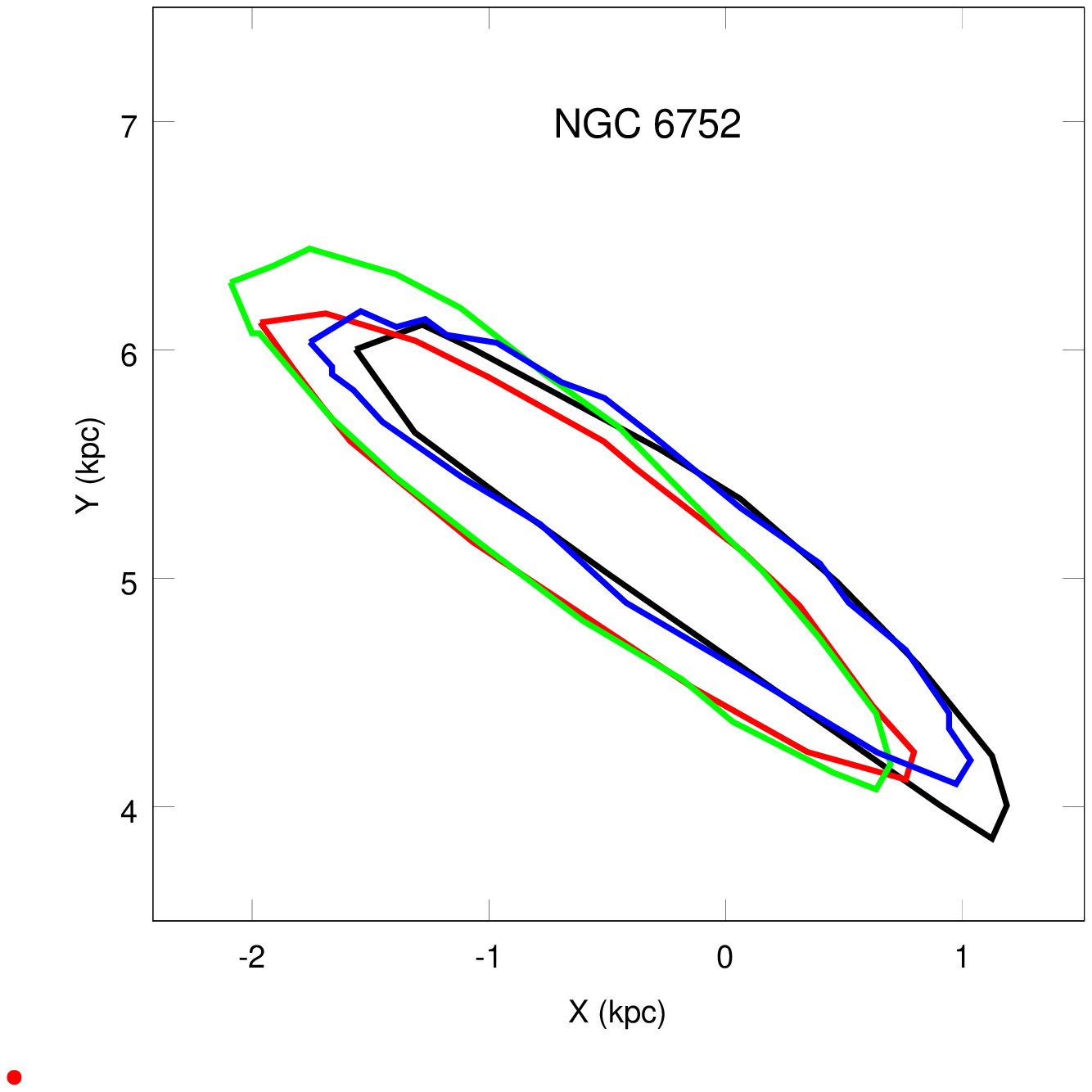} &
\includegraphics[bb=30 42 490 469,height=7.5cm,clip=]{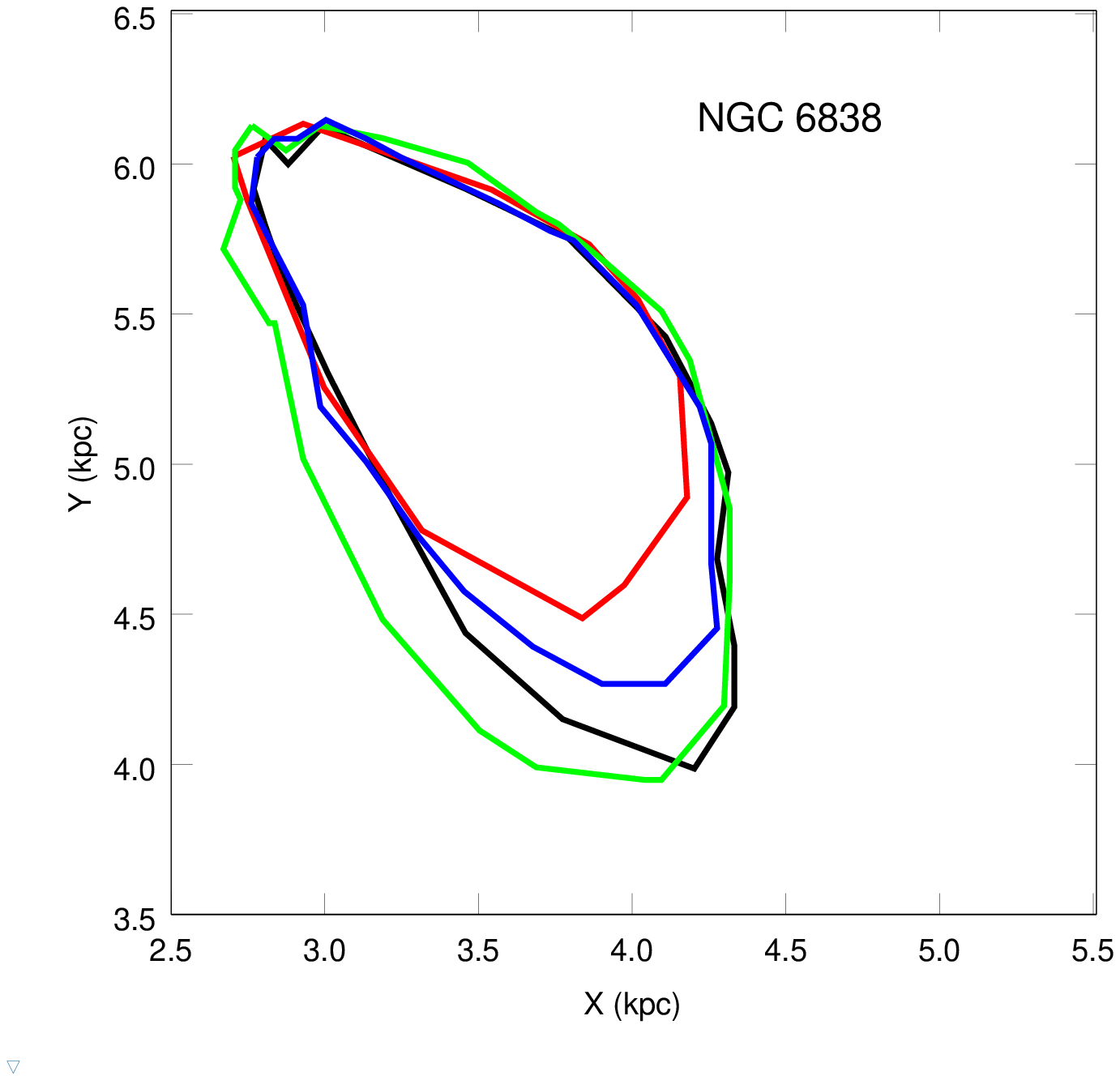} \\
\end{tabular}
\end{center}
\caption{Present-day position of the 90$\%$ contour levels for the 6 GCs and 4 potentials (DA--black, FE--red, FL--green, PA--blue).  The Sun (red circle) is at x=0 kpc, y=8 kpc. }

\end{figure*}

\begin{figure}
\includegraphics[bb= 20 40 500 469,width=94mm]{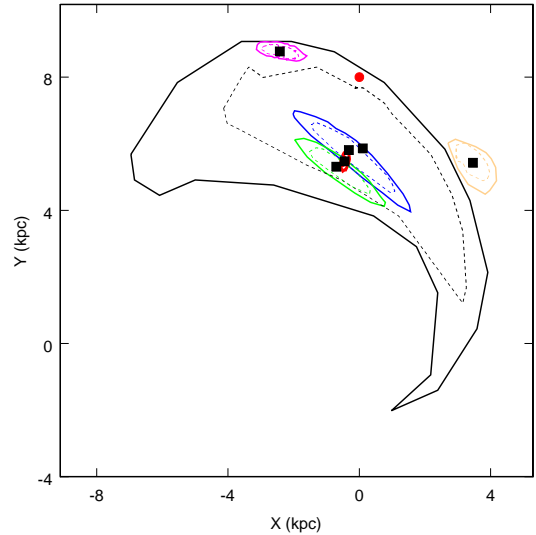}
\caption{Present-day Galactic plane location of the 90 $\%$ (solid line), and 68 $\%$ (dashed line) contour levels for probability of impact, for NGC104(blue), NGC1851 (black), NGC3201 (pink), NGC6397 (red), NGC6752 (green), and NGC6838 (orange). Nominal impact positions are black squares.
The Sun (red circle) is at x = 0 kpc, y = 8 kpc, and only the FE potential is used here.}
\end{figure}

\subsection{Impact points of NGC 3201, NGC 6397, and NGC 6838.}

The six areas and subtended fields of view from Earth encompassed by the contours in Figure 7 are smallest for NGC 3201, NGC 6397 
and NGC 6838. As a result, these offer the best prospects for identifying 
remnants of crossings. For NGC 3201, the area is enclosed within a radius of 
 7\r{} on the sky, and a distance range of $\pm $22\% from the nominal impact.  For NGC 6397, these radii are 
 3\r{}, with distance  $\pm $12\% and for NGC 6838, they are 6\r{}, and $\pm $20\%.

We searched the CDS Simbad database\footnote{http://cdsweb.u-strasbg.fr/, status March 2008} centred on regions of sky corresponding to these areas, and restricted the search to object 
types ``Star cluster, Cl*", ``Open Galactic cluster, OpC", and ``Association of 
stars, As*" as defined in the database. This leaves 15, 33 and 42 star groups respectively, for NGC 3201, NGC 6387, and NGC 6838, and these are listed in Tables 5, 6 and 7.  
\bigskip

\begin{table*}
\centering
\caption{Star groups within the directions of NGC 3201 90\% contours 
in Figure 7}
\begin{tabular}
{|p{77pt}|p{17pt}|p{80pt}|p{24pt}}
\hline
Identifier& 
Type& 
Basis for exclusion& 
Refs\\
\hline
C 0820-360 & 
Cl* & 
A, D& 
1 \\

C 0815-369 & 
Cl* & 
A& 
2 \\
C 0813-306 & 
Cl* & 
A, D& 
1 \\
NGC 2546 & 
Cl* & 
A, D&  
1 \\
ESO 430-18 & 
Cl* & 
A& 
1 \\
$[$KPR2005$]$ 45 & 
Cl* & 
A& 
3 \\
C 0812-362 & 
Cl* & 
A, D& 
1 \\
 C 0810-324 & 
OpC & 
A, D& 
 1 \\
C 0812-318 & 
OpC & 
A, D& 
4\\
NGC 2588 & 
OpC & 
A, D& 
5 \\
\hline
\end{tabular}
\\Potential candidates are C 0805-322, [DBS2003] 16, C 0807-343, C 0809-318, DSH J0807.1-3603, pending distance and age data\\Refs: 1, Kharchenko, 2005a; 2, Lindoff, 1968; 3, Kharchenko et al., 2005b; 4,
 Giorgi et al., 2007; 5, Baume et al., 2004.\\
\end{table*}

\begin{table*}
\centering
\caption{Star groups within the directions of NGC 6397 90\% contours 
in Figure 7}
\begin{tabular}
{|p{87pt}|p{26pt}|p{80pt}|p{24pt}}
\hline
Identifier& 
Type& 
Basis for exclusion& 
Refs \\
\hline
C 1715-382 & 
Cl* & 
D& 
1 \\
$[$BDS2003$]$ 97 & 
Cl* & 
D& 
2 \\
$[$BDS2003$]$ 98 & 
Cl* & 
D& 
2 \\
$[$BDS2003$]$  164& 
Cl* & 
D& 
2\\
$[$BDS2003$]$ 99 & 
Cl* & 
D& 
2\\
C 1714-355 & 
Cl* & 
A, D& 
4\\
$[$DBS2003$]$ 179 & 
Cl* & 
D& 
5 \\
NGC6318 & 
OpC & 
A& 
3 \\
\hline
\end{tabular}
\\Potential candidates are [DBS2003] 120; [DBS2003] 121; [DBS2003] 122; [DBS2003] 123;MFSW VI; MFSW V; MFSW IV; C 1717 358; MFSW III; MFSW II; MFSW I; DSH J1715.7-3843; [BDB2003]G351.61+.17; [BDS2003] 119; [DBS2003] 118; C 1712-393; C 1727-370; Cl VDBH 214; C 1720-378; C1721-389; C 1728-368; DSH J1713.2-3942; C 1715-387; [DBS2003] 166; [DBS2003] 165, pending distance and age data\\Refs: 1, Piatti \& Clari\'{a}, 2002; 2, Bica et al., 2003; 3, Piatti et al., 2000; 4, Kharchenko et al., 2005b; 5,
 Borissova et al, 2005\\
\end{table*}

\begin{table*}
\centering
\caption{Star groups within the directions of NGC 6838 90\% contours in Figure 7}
\begin{tabular}
{|p{81pt}|p{26pt}|p{80pt}|p{24pt}|}
\hline
Identifier& 
Type& 
Basis for exclusion& 
Refs \\
\hline
$[$KPR2005$]$ 104 & 
Cl* & 
A, D& 
1\\
C 1915 +194 & 
Cl* & 
A& 
2 \\
$[$BDS2003$]$ 155 & 
Cl* & 
D& 
3 \\
$[$BDS2003$]$ 153 & 
Cl* & 
D & 
3 \\
$[$BDS2003$]$ 154 & 
Cl* & 
D& 
3 \\
$[$BDS2003$]$ 146 & 
Cl* & 
D& 
3 \\
$[$BDS2003$]$ 152 & 
Cl* & 
D& 
3 \\
$[$BDS2003$]$ 150 & 
Cl* & 
D& 
3 \\
$[$BDS2003$]$ 151 & 
Cl* & 
D& 
3 \\
$[$BDS2003$]$ 145 & 
Cl* & 
D& 
3 \\
OB cluster in W51 & 
Cl* & 
D& 
3 \\
$[$BDS2003$]$ 148 & 
Cl* & 
D& 
3 \\
$[$BDS2003$]$ 147 & 
Cl* & 
D& 
3 \\
$[$BDS2003$]$ 144 & 
Cl* & 
D & 
3 \\
$[$BDS2003$]$ 143 & 
Cl* & 
D& 
3 \\
$[$BDS2003$]$ 142 & 
Cl* & 
D& 
3 \\
$[$BDS2003$]$ 141 & 
Cl* & 
D& 
3 \\
$[$BDS2003$]$ 140 & 
Cl* & 
D& 
3 \\
NGC6793 & 
Cl* & 
A, D& 
1 \\
$[$BDS2003$]$ 139 & 
Cl* & 
D& 
3 \\
$[$BDS2003$]$ 138 & 
Cl* & 
D& 
3 \\
DSH J1930.2+1832 & 
OpC & 
A D& 
4 \\
%\hline
DSH J1933.9+1831 & 
OpC & 
A, D& 
4 \\
NGC6802 & 
OpC & 
A, D& 
5\\
C 1942+174 & 
OpC & 
A, D& 
1\\
\hline
\end{tabular}
\\Potential candidates are [BDS2003] 12; [BDS2003] 156; C 1926+173; DSH J1926.0+1945; [BDS2003]157; [BDS2003] 13; C 1926+147; DSH J1942.3+1939; C 1916+156; DSH J1942.8+1530; C 1940+210; DSH J1912.0+1716; C 1922+136; DSH J1937.3+1841; Cl Czernik 40; DSH J1925.2+1356; Cl Alessi 57, pending distance and age data\\ Refs: 1, Kharchenko et al., 2005b; 2, Carraro et al., 2006; 3, Bica, et al., 2003; 4, Kromberg et al., 
2006;  5, Netopil et al., 2007.\\
\end{table*}

We searched the relevant literature for details of age and distance.  Where this information is available, we noted for each star group in Tables 5, 6, and 7, the parameter that is incompatible, age (A) by a factor of at least two, and distance (D) outside the bounds given above.  We conclude that of a total of fifteen star groups that lie in the direction of NGC3201's possible impact remnants, ten are definitely not remnants, on the grounds of incompatible age or distance (Table 5).  In the remaining five cases, it is possible that the groupings could result from the impact, pending further work on determining the properties (age, distance) of the associations.  Examining the corresponding data for NGC6397 (Table 6), out of thirty-three candidates, eight can be excluded.  The corresponding data for NGC6838 (Table 7) indicates that twenty five out of forty two candidates,  can be excluded.
This means that we are able to discount some GCs, and identify potential candidates, pending availability of further data.

\section{Could Super Star Clusters be remnants?}

\begin{figure}
\includegraphics[width=89mm,clip]{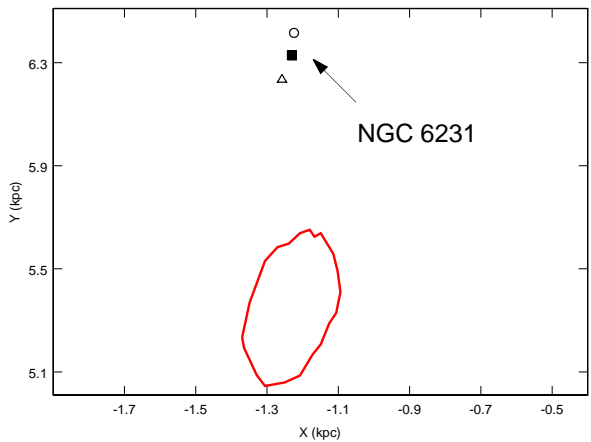}
\caption{Position of open cluster NGC 6231 and of the 90$\%$ level of NGC 6397 impacts (FE potential), at the time of GC crossing.  The triangle and circle represent the points corresponding to today's distance estimate $\pm 3\sigma$, respectively.}
\end{figure}

An alternative approach to linking star formation to GC disk crossings is to 
work in reverse, looking for evidence of unusual star formation, and asking whether this could be due to a GC disk crossing.  For this we have concentrated on Super Star Clusters (SSCs).  Portegies Zwart (2004) identifies five such SSCs: Arches, Quintuplet, NGC3603, Westerlund 1, and Westerlund 2. Their distance from the 
Galactic centre and age appear in Table 8.

\subsection{Monte Carlo impact analysis}

We first look for progenitors amongst our sample of fifty-four GCs as we have already determined their crossings.  It is possible to eliminate some of these, on the basis of the time elapsed since the last crossing.  The time $t$ since a crossing is given by $t\, = \,t_{C} \, + \,t_{SF} \, + \,t_{A} $ where $t_{C} \,$is the time interval between disk crossing 
and the onset of star formation, $t_{SF} $ is the duration of star formation, and $t_{A} $ is the age of the 
cluster. The first time ($t_{C} \,$) is not well known, so three cases are 
considered here. Firstly, it is possible that this time is small compared to 
the age of the cluster, so that we can neglect it.  Secondly, we adopt 
the value of 15Myr given by L\'{e}pine \& Duvert (1994) for the case of star 
formation after a High Velocity Cloud impacts the disk. Finally, we take the 
value of 30Myr suggested by Wallin et al. (1996). For 
the star formation time, we adopt a value of $t_{SF}  $=2 $\times$10$^{{\rm 5}}$ years, 
suggested by McKee and Tan (2002), for massive stars ($M > \,8\,M_{ \odot}  
$). 
Taking the value $t_{C} \,$=30 Myr, and an average SSC age of $t_{A} \,$=5 Myr, we are able to identify 24 GCs as not being SSC progenitors, because the time since impact is too long compared with 35Myr, using the FE potential.   As a result, the list of fifty-four GCs to consider for Monte Carlo analysis is reduced to 30 candidates.  A Monte Carlo analysis of these 30 candidates, similar to those above, using the FE potential, then shows further that the SSCs could not have been produced by any of these 30 GCs, with the exception of NGC362, 5897, 6093, 6316, 6522, 6584, and 6779, whose impact footprint is compatible with both the Arches and Quintuplet.  Hence here too, some GCs can be discounted, while others may be potential progenitors.

\subsection{Escape velocity analysis}
Having considered the fifty-four GCs, we then broaden our scope to all one hundred and fifty-five GCs in the Harris catalogue, and note that we can also screen for candidates by using a weaker method but which does not require knowledge of any proper motions or radial velocities.  As the position of the GC at the time of impact is the same as the position 
of the SSC's ancestral gas cloud, it is possible to calculate the 
minimum velocity $V$ required for the GC to cover the distance from that position, to 
its present-day position within a time  $t \,$.  The escape velocity at the Galactocentric distances ($D$) in Table 8 can then be 
calculated from the potential, using the relation $V_{esc} \, = \,\sqrt 
{2.{\left| {\Phi (D)} \right|}} $ (Binney and Tremaine, 1994). If $V$$ > $
$V_{esc {\rm} {\rm ,}{\rm} }$then the GC's last disk crossing cannot be the 
event that triggered the formation of the SSC.

We made the comparison for all GCs in the Harris catalogue, all five 
Super Clusters, the FE potential, and three times $t_{C} \,$ discussed 
above. Table 9 summarises the analysis.  We can identify the GCs that are definitely not progenitors of the SSCs.  As expected and evident in Table 9, the case with the smallest set of excluded GCs is that where $t_{C} \,$=30 Myr, and Table 10 lists the definitely excluded GCs.  On the other hand, if star formation as a result of disk crossing is prompt ( $t_{C} \,$$\sim$0) then we can exclude \textit{all} GC crossings as triggering the formation of both NGC3603 and Westerlund 2.  If this is generally the case, it may indicate that SSCs are not generated by GC disk crossings.

\begin{table}
\centering
\caption{Distance from Galactic centre, and ages of  Super Star Clusters}

\begin{tabular}
{|p{53pt}|p{51pt}|p{38pt}|p{21pt}|}
\hline
& 
Distance from Galactic centre \par $D$ (kpc)& 
Age \par Myr& 
Refs. 
 \\
\hline
Arches& 
$\sim {\rm 0}{\rm .}{\rm 1}$
& 
$ {\rm 2}{\rm .}{\rm 5}$$\pm0.5$
& 
1 \\
Quintuplet& 
$\sim {\rm 0}{\rm .}{\rm 05}$
&
$ {\rm 4}$$\pm1$
& 
2 \\
NGC3603& 
8$\pm0.4$& 
$\sim {\rm 1}$
& 
3, 4 \\
Westerlund1& 
$\sim {\rm 9}{\rm .}{\rm 4}$
& 
1-10& 
5 \\
Westerlund2& 
9.8$\pm1.7$& 
$\sim {\rm 2.5}$
& 
6 \\
\hline
\end{tabular}
\\ Refs: 1, Figer et al., 2002; 2, Figer, McLean, Morris, 1999; 3, 
Stolte et al., 2004; 4, Stolte et al., 2006; 5, Vrba et al., 2000; 6, Rauw et 
al., 2007.  \\
\end{table}

 \begin{table}
\centering
\caption{ Percentage of all Globular Clusters from the Harris catalogue 
that could have spawned the Super Clusters, based on their velocity being 
less than the escape velocity.}
\begin{tabular}
{|p{55pt}|p{46pt}|p{50pt}|p{50pt}|}
\hline
& 
$t_{C} \,$ = 0 Myr& 
$t_{C} \,$ = 15Myr& 
$t_{C} \,$ = 30Myr  \\
\hline
Arches& 
20& 
76& 
93 \\
%\hline
Quintuplet& 
34& 
80& 
93 \\
NGC 3603& 
0& 
16& 
75 \\
%\hline
Westerlund 1& 
1& 
60& 
82 \\
%\hline
Westerlund 2& 
0& 
17& 
73 \\
\hline
\end{tabular}
\end{table}

\begin{table*}
\centering
\caption{List of globular Clusters that are definitely not progenitors of the five SSCs for the case $t_{C}\,$=30Myr.}
\begin{tabular}
{|p{58pt}|p{83pt}|p{75pt}|p{85pt}|p{71pt}|p{90pt}|}
\hline
GC& 
SSC& 
GC& 
SSC& 
GC& 
SSC \\
\hline
AM1 & 
A, Q, N, W1, W2& 
NGC5634 & 
N, W1, W2& 
NGC7492& 
N, W1, W2 \\
%\hline
AM4 & 
N, W1, W2& 
NGC5694& 
A, Q, N, W1, W2& 
Pal1 & 
N, W2 \\
%\hline
Arp2 & 
N, W1, W2& 
NGC5824& 
N, W1, W2& 
Pal2& 
A, Q, N, W1, W2 \\
%\hline
Eridanus & 
A, Q, N, W1, W2& 
NGC6205& 
W2& 
Pal3& 
A, Q, N, W1, W2 \\
%\hline
ESOSC06& 
W1& 
NGC6229 & 
A, Q, N, W1, W2& 
Pal4 & 
A, Q, N, W1, W2 \\
%\hline
IC1257 & 
N, W1, W2& 
NGC6341 & 
N, W2& 
Pal5 & 
N, W1, W2 \\
%\hline
IC4419& 
W1& 
NGC6426& 
N& 
Pal11 & 
N, W2 \\
%\hline
NGC1261& 
W1& 
NGC6715 & 
N, W1, W2& 
Pal12 & 
N, W2 \\
%\hline
NGC1904 & 
W2& 
NGC6779& 
N, W2& 
Pal13 & 
N, W1, W2 \\
%\hline
NGC2419 & 
A, Q, N, W1, W2& 
NGC6864& 
N, W2& 
Pal14& 
A, Q, N, W1, W2 \\
%\hline
NGC4147 & 
N, W1, W2& 
NGC6934 & 
N, W2& 
Pal15& 
A, Q, N, W1, W2 \\
%\hline
NGC5024 & 
N, W2& 
NGC6981& 
N, W2& 
Pyxis& 
A, Q, N, W1, W2 \\
%\hline
NGC5053 & 
N, W2& 
NGC7006& 
A, Q, N, W1, W2& 
Rup106& 
W1 \\
%\hline
NGC5272 & 
W2& 
NGC7078& 
N, W2& 
Terzan7& 
N, W1, W2 \\
%\hline
NGC5466 & 
N, W2& 
NGC7089 & 
N, W2& 
Terzan8& 
N, W1, W2 \\
\hline
\end{tabular}
\\ A=Arches; Q=Quintuplet; N=NGC3603; W1=Westerlund1; W2=Westerlund2\\
\end{table*}

\section{CONCLUSIONS}

Globular Cluster interaction with a Galactic disk has been studied in detail from the viewpoint of the GC, see for example Gnedin, Lee \& Ostriker (1999).  On the other hand, the effect on the disk has received scant attention (Brosche et al. 1991, Wallin et al. 1996, Levy 2000).  

The suggestions in these papers is that GC disk 
crossings could cause star formation. We have 
therefore explored the process in more detail. We prescribe the Galactic 
potential, and allow the GC to move within it.  We also model the relative movement of a cloud of disk material on a circular orbit, as it encounters the GC crossing the disk, and show that focussing of disk material will occur for several of the GCs.  The compression is on a scale of $\sim$ 10pc, as suggested by Wallin et al.   The extent of compression increases with GC mass and time spent by the GC in the cloud.  Even if no gas is present at the time of interaction, the concentrated disk material can subsequently attract gas, leading possibly to star formation.

Should gas be immediately available in the impact region, a potentially more effective mechanism exists, proposed by Levy (2000), based on shock wave compression by the GC transiting through the disk.  The star forming region in this case is estimated to be of kpc size.

We have then examined whether there is any 
observational evidence for star formation as a 
result of a GC disk crossing. The uncertainties 
in the parameters in our subset of fifty-four GCs, 
combined with the uncertainty in the potential 
model are such that the impact regions are 
generally too extensive, subtending large angles 
on the sky. For some nearby GC crossings this is not the 
case, however, and for NGC 
3201, NGC 6397, and NGC 6838, the 
subtended angles were smaller, as determined by 
Monte Carlo simulations. Using the CDS we 
identified a number of candidate star clusters 
and associations within the impact regions. We 
were also able to exclude a number of clusters 
and associations within the 90\% contour on the 
grounds of age or distance. The only prior 
candidate for GC disk crossing star formation, 
the open cluster NGC 6231, lies well outside the 
90\% contour of NGC 6397 impacts, in 
contradiction of the suggestion by Rees \& 
Cudworth in Wright (2004).

We then reversed the approach, and considered 
whether any of the Galaxy's five most prominent 
Star Super Clusters (Arches, Quintuplet, NGC 
3603, Westerlund 1 and Westerlund 2) could have 
resulted from an impact by any of the sample of 
fifty-four GCs, using age arguments, and the Monte 
Carlo technique applied above. From this we 
found that only 7 GCs are candidates to provide 
impact site progenitors for the Arches and 
Quintuplet SSCs, with none for the other SSCs. 
As a final test, we considered all 155 GCs in 
the Harris catalogue and determined which SSCs 
could conceivably have resulted from a GC disk 
crossing. We used the criterion that the minimum
speed to travel from a prior position of the 
SSCs to the current GC position should not 
exceed the Galactic escape velocity. This 
eliminated many of the larger sample of GCs, 
depending on the time elapsed between disk 
crossing and the onset of star formation. If the 
time between the disk crossing and the onset of 
star formation is limited to a few Myr, it is 
possible to exclude any of the Galactic GC disk 
crossings as a potential cause of NGC 3603 and 
Westerlund 2 formation.

\section*{Acknowledgments}

This research used the Smithsonian/NASA ADS, NASA's Heasarc coordinate converter, and the CDS at Strasbourg.  DVP acknowledges the support of an STFC grant.  Mike Fellhauer, Ignacio Ferreras, Chris Flynn,  Esko Gardner, Jason Kalirai, Barbara Pichardo, and Kinwah Wu helped us greatly by sharing their knowledge of Galaxy modelling.  We are also grateful to the referee who helped us clarify some of the aspects of this work, thus improving the paper.

\end{document}